\documentclass[fleqn,usenatbib]{mnras}

\usepackage{newtxtext,newtxmath}

\usepackage[T1]{fontenc}

\usepackage[normalem]{ulem} 
\DeclareRobustCommand{\VAN}[3]{#2}
\let\VANthebibliography\thebibliography
\def\thebibliography{\DeclareRobustCommand{\VAN}[3]{##3}\VANthebibliography}

\usepackage{graphicx}	
\usepackage{amsmath}	

\title[Distinct regimes of LRDs and LBDs]{Beyond orientation: Evidence for distinct physical regimes among Little Red Dots and Little Blue Dots}

\author[L. Barrufet]{L. Barrufet$^{1}$\thanks{E-mail: lbarrufe@roe.ac.uk},
J. S. Dunlop$^{1}$, 
M. L. Hamadouche$^{2}$, 
M. Mezcua$^{3}$, 
D. Herrero-Carrion$^{4,5}$, 
M. Santos-Lleo$^{6}$, 
\newauthor
R. Gottumukkala$^{7, 8}$, 
D. Spinoso $^{9}$, 
C. T. Donnan  $^{10}$
\\
$^{1}$ Institute for Astronomy, University of Edinburgh, Royal Observatory, Edinburgh, EH9 3HJ, UK 
\\
$^{2}$ Department of Astronomy, University of Massachusetts, Amherst, MA 01003, USA
\\
$^{3}$ Institute of Space Sciences (ICE, CSIC), Campus UAB, Carrer de
Can Magrans, s/n, 08193 Barcelona, Spain
\\
$^{4}$ Donostia International Physics Centre (DIPC), Paseo Manuel de Lardizabal 4, 20018 Donostia-San Sebastian, Spain
\\
$^{5}$ University of the Basque Country UPV/EHU, Department of Physics, Barrio Sarriena s/n. 48940 Leioa, Bizkaia, Spain
\\
$^{6}$ Formerly, European Space Astronomy Centre (ESAC), ESA, Camino Bajo del Castillo s/n, 28692 Villanueva de la Canada, Madrid, Spain
\\
$^{7}$ Cosmic Dawn Center (DAWN)
\\
$^{8}$ Niels Bohr Institute, University of Copenhagen, Jagtvej 128, DK-2200 Copenhagen North, Denmark
\\
$^{9}$ Dipartimento di Fisica “G. Occhialini”, Università degli Studi di Milano-Bicocca, Piazza della Scienza 3, I-20126 Milano, Italy
\\
$^{10}$ NSF’s National Optical-Infrared Astronomy Research Laboratory, 950 N. Cherry Ave., Tucson, AZ 85719, US
}
\date{August 2026}

\pubyear{\the\year{2026}}

\begin{document}
\label{firstpage}
\pagerange{\pageref{firstpage}--\pageref{lastpage}}
\maketitle

\begin{abstract}

\noindent Little Red Dots (LRDs) and Little Blue Dots (LBDs) may represent the same population of compact active galactic nuclei (AGN) observed along different lines of sight. We test this scenario using a spectroscopic sample from the DAWN {\it JWST} Archive, selected based on common criteria for H$\alpha$ equivalent width, UV continuum slope, and compactness. 
We use the optical continuum slope to distinguish between samples of 89 LRDs and 191 LBDs spanning $1 \lesssim z \lesssim 7.5$, and compare their continuum properties, H$\alpha$ emission, and Balmer decrements. Consistent with previous studies, we find that LRDs exhibit broader H$\alpha$ profiles than LBDs systematically (with median FWHM values of $2319^{+71}_{-64}$ and $1424^{+56}_{-43}~\mathrm{km\,s^{-1}}$, respectively). We also confirm that LRDs show larger Balmer decrements, with median $\log_{10}(F_{\rm H\alpha}/F_{\rm H\beta})=1.01\pm0.03$, compared with $0.47\pm0.01$ for LBDs. Line-of-sight effects could explain both results. However, most significantly, we find that the LRD fraction increases strongly with H$\alpha$ line luminosity: LRDs are approximately six times more luminous in H$\alpha$ than LBDs. This robust finding is much harder to explain through unification with LBDs by orientation. Despite their higher line luminosities, LRDs have a moderately lower median H$\alpha$ equivalent width than LBDs, consistent with more of the raw emission being reprocessed into the rest-frame optical continuum in LRDs. These results, coupled with the systematically lower [O III]/H$\beta$ ratios found in LRDs and the decline of H$\alpha$ equivalent width towards the reddest optical slopes, disfavour a simple orientation-based link between LRDs and LBDs, and are instead more consistent with the predictions of gas-cocoon models in which increasing gas column density explains the apparent transition from LBDs to LRDs.

\end{abstract}




\section{Introduction}
\label{sec:introduction}

Little Red Dots (LRDs) have emerged as one of the most unexpected discoveries enabled by the {\it James Webb Space Telescope} (\textit{JWST}). Initially identified in photometric samples of high-redshift sources, their bright rest-frame optical emission was interpreted as evidence for unexpectedly massive galaxies at early times \citep{Labbe2023}. Subsequent observations showed that strong emission lines and active galactic nucleus (AGN) emission contribute significantly to their rest-frame optical emission. At the same time, spectroscopy revealed compact sources with a blue UV continuum slope and a red rest-frame optical continuum, characterised by a distinctive V-shaped spectral energy distribution \citep{Matthee2024}. Larger photometric and spectroscopic samples have since shown that LRDs are common at $z\gtrsim3$, with relatively little evolution in their number density at $z\gtrsim4$ \citep{PerezGonzalez2024}. However, recent studies suggest more complex evolution at lower redshifts \citep{Kapoor2026, inayoshi2024_numberDensities, Loiacono2026}, while LRDs appear to span a broad, possibly continuous range of continuum and emission-line properties \citep{deGraaff2025DJA, Barro2026photsel, Rinaldi2026, PerezGonzalez2026, matthee2026_photometricSequence}.

Despite this rapid observational progress, the physical nature of LRDs remains unclear. Their compact morphologies, broad Balmer emission, and high-ionisation lines favour the presence of accreting black holes \citep{Matthee2024, Greene2024, deGraaff2025DJA, Maiolino2025, Papovich2026}. However, most remain individually undetected in X-rays and appear substantially X-ray weaker than expected from standard AGN spectral energy distributions \citep{Ananna2024, Maiolino2025}. Their weak or absent thermal dust emission challenges conventional dust-obscured AGN models \citep{Casey2024, Akins2025}, while their broad emission lines, combined with weak X-ray and infrared emission, are difficult to reconcile with a scaled-down interpretation of classical quasars (QSOs) as a whole.  Nevertheless, growing spectroscopic evidence favours a compact central engine in many LRDs, and some show signatures of evolving towards more typical AGN and QSOs \citep{fu2026_lrd_into_qsos,ji2026_holesbhstar,korber2026_lrdoutflow}. The origin of their unusual continua and broad emission lines, and the relative contributions of stellar and AGN emission, therefore remain under debate \citep{PerezGonzalez2024, HerreroCarrion2026, PerezGonzalez2026, Baggen2024, Scholtz2026}.

Several recent models instead invoke a compact accreting source surrounded by a dense gaseous environment to explain the unusual combination of LRD properties. In these scenarios, surrounding material can absorb and reprocess the intrinsic emission, potentially suppressing the escaping X-rays while contributing to the optical continuum and Balmer-line emission. Proposed interpretations include rapidly accreting black holes enclosed by dense ionised cocoons \citep{Rusakov2026, Sneppen2026_cocoonmodel}, self-obscured super-Eddington systems \citep{Inayoshi2025, Madau2026}, and ``black-hole stars'', in which a compact central source is embedded within a gas-dominated photosphere \citep{Naidu2025, Sun2026}. Despite their differences, these models share the idea that the observed properties of LRDs are strongly influenced by the physical conditions of the gas surrounding the central engine. 

The recent identification of compact sources with similar emission-line properties but blue rest-frame optical continua has further broadened the diversity of this population. Commonly referred to as Little Blue Dots (LBDs), these sources were first defined by \citet{Brazzini2026}.  Like LRDs, LBDs exhibit X-ray weakness and broad Balmer emission, but their spectra lack the characteristic V-shaped continuum. The relationship between the two populations remains unclear: some observational studies suggest a continuous transition between LBDs and LRDs \citep{Asada2026}, whereas others interpret them as physically distinct classes \citep{Brazzini2026}. 

Two broad scenarios have been proposed for the connection between LBDs and LRDs. In the orientation-based model, LRDs and LBDs are SMBHs accreting at super-Eddington rates and represent the same underlying population viewed along different lines of sight, analogous to the classical AGN unification model \citep{Madau2026, Madau2026wings}. Alternatively, models proposed by \citet{Sneppen2026_cocoonmodel, Sneppen2026_LBDLRDs} place a gas cocoon surrounding the accreting source, with the different spectral properties arising from changes in the physical conditions of the gas around the central engine. In this scenario, variations in the gas column density are the primary driver of the observed spectral differences, linking LBDs and LRDs as different manifestations of the same underlying population. These scenarios make different predictions for the observable properties of LBDs and LRDs, but have not yet been tested systematically using a homogeneous observational sample. 

In this work, we construct and compare homogeneous samples of LRDs and LBDs to investigate whether their observed properties are linked primarily by orientation or instead reflect intrinsic physical differences. We develop a new selection based on the properties known to characterise these compact sources, combining strong H$\alpha$ emission, a blue UV continuum, and compact morphology, and then distinguish LRDs from LBDs according to their rest-frame optical continua. Section~\ref {sec: observations} describes the observations, while Section~\ref{sec: selection} details the sample selection and analysis methodology. We present our results in Section~\ref{sec: results}, discuss their implications in Section~\ref{Section: Discussion}, and summarise our conclusions in Section~\ref{sec:summary_conclusions}.

\section{Observations}
\label{sec: observations}

In this section, we describe the \textit{JWST}/NIRSpec spectroscopy and ancillary \textit{JWST}/NIRCam imaging used to construct and characterise the LRD and LBD samples, together with the archival X-ray data used to search for counterparts.

\subsection{NIRSpec spectroscopic data}
\label{sec: spectra observations}

We use the public \textit{JWST}/NIRSpec spectroscopy compiled in version~4.4 of the DAWN JWST Archive (DJA; \citealt{Valentino2025}), which contains approximately 86,000 spectra from a wide range of observing programmes and extragalactic fields. The archive primarily comprises low-resolution PRISM observations obtained with the NIRSpec micro-shutter assembly, complemented by medium- and high-resolution spectroscopy where available. The spectra were reduced uniformly with \texttt{msaexp}, including background subtraction, optimal one-dimensional extraction, wavelength and flux calibration, and corrections for wavelength-dependent slit losses \citep{Heintz2024_primal, Pollock2026}.

We apply the same initial selection to LRDs and LBDs. To ensure reliable source identification and redshift measurements, we retain only spectra with \texttt{grade=3}, corresponding to robust spectroscopic redshifts based on the DJA quality assessment. This reduces the parent sample to 37,528 spectra. We adopt the DJA spectroscopic redshift, \texttt{zline}, and use the reduced one-dimensional spectra to measure the H$\alpha$ equivalent width and UV continuum slope for each source. These measurements are combined with the compactness derived from the NIRCam imaging described below to construct the final sample (Section~\ref{sec: selection}).

\subsection{Photometric data}
\label{sec: photometry}

Although the LRD and LBD classifications are primarily based on spectroscopic properties, ancillary imaging is required to quantify the rest-optical compactness of the sources. We retrieve the corresponding \textit{JWST}/NIRCam F444W mosaics from the DJA. This filter provides the most widely available long-wavelength NIRCam coverage across the heterogeneous fields included in the spectroscopic archive.

We determine the source centroid in the F444W image and perform aperture photometry within circular apertures of $0.1^{\prime\prime}$ and $0.2^{\prime\prime}$ radii, including the associated photometric uncertainties. From these measurements, we calculate the compactness parameter
\[
\mathcal{C}_{\rm F444W}
=
\frac{F_{\rm F444W}(0.2^{\prime\prime})}
     {F_{\rm F444W}(0.1^{\prime\prime})},
\]
following the approach of \citet{deGraaff2025RUBIES}. Lower values indicate that a larger fraction of the light is concentrated near the source centre. The compactness criterion applied to the sample is described in Section~\ref{sec: selection}.

We additionally measure F444W fluxes within $0.4^{\prime\prime}$-radius apertures and correct them to total fluxes using the NIRCam curves of growth, from which we additionally derive total AB magnitudes. At redshift $z = 6$ ($z = 2$), this corresponds to $\sim 2.4$ ($\sim3.4$) kpc, which is substantially larger than the expected radii of LRDs at these redshifts \citep[$\lesssim 1$ kpc, see e.g.,][]{cloonan2026}. These measurements are used to characterise the photometric and structural properties of the final LRD and LBD samples. 

To characterise the X-ray properties of the sample, we use archival data from the two X-ray catalogues providing the deepest per-pointing sensitivity over the fields covered by our sample: 5XMM-DR15 \citep{Webb2026} and the Chandra Source Catalogue (CSC 2.1; \citealt{Evans2024}). We queried the XMM-Newton Science Archive and the CSC Quick Search interfaces for X-ray sources and available coverage at the positions of the sample galaxies. The cross-matching procedure and the identification of reliable X-ray counterparts are described in Section~\ref{sec: selection}.

\section {Little (Red and Blue) dots Methodology and Selection} 
\label{sec: selection} 

To construct a large and homogeneous sample of compact AGN-like sources, we base our selection on three observables that capture the main phenomenological properties of LRDs and LBDs: strong broad H$\alpha$ emission, a blue UV continuum, and compact morphology. Broad-line selections are commonly used to identify AGN \citep{Maiolino2025} 
and have also been applied in the context of LRDs. More recently, \cite{Rusakov2026} used broad H$\alpha$ emission, defined as \hbox{FWHM $>1000~\mathrm{km\,s^{-1}}$}, to select LRDs. In their sample of twelve sources, however, only three are defined as LRDs, while the remaining objects are classified as LRD-like AGN. These results illustrate that broad-line selection alone does not uniquely identify LRDs. Similarly, \citet{Hviding2025} distinguish a wider population of broad-Balmer AGN from spectroscopic LRDs, defining the latter as sources that additionally exhibit a V-shaped continuum and a dominant rest-optical point-source component. 

However, FWHM measurements are particularly sensitive to spectral resolution and line-profile modelling, both of which are important limitations for the predominantly low-resolution JWST/NIRSpec spectra available for large samples.

For these reasons, we use the H$\alpha$ equivalent width as a more robust indicator of strong line emission. Large H$\alpha$ equivalent widths are a characteristic property of LRDs \citep[e.g., see][] {deGraaff2025DJA}, consistent with their high H$\alpha$ luminosities and faint underlying continua. 
This allows us to identify sources with prominent H$\alpha$ emission without requiring a resolved broad-line component.

A second defining property of LRDs is their blue UV slope. This criterion was already present in the first works defining LRDs, such as \citet{Matthee2024}, which required a blue UV continuum. However, the exact threshold varies between studies, from approximately $\beta_{\rm UV}<-0.5$ to $\beta_{\rm UV}<-0.2$. 
Finally, compactness is one of the most distinctive properties of LRDs and is also expected for AGN-dominated sources. Combining these three observables allows us to identify a broad parent population of compact sources with strong and broad H$\alpha$ emission and blue UV continua, without imposing the red or blue rest-frame optical continuum that distinguishes LRDs from LBDs. The LRD and LBD populations are then distinguished according to their rest-frame optical continuum properties, as described below.

\subsection{Measuring $\beta_{\mathrm{UV}}$, $\beta_{\mathrm{opt}}$ and H$\alpha$ emission line properties}\label{sec:beta_slope_measurements}

The UV slope was measured over the rest-frame wavelength interval
\(1300 \leq \lambda_{\rm rest} < 3646\,\text{\AA}\), where
\(3646\,\text{\AA}\) corresponds to the Balmer limit. Before fitting,
regions of \(\pm30\,\text{\AA}\) around prominent UV spectral features
were excluded. These comprise O\,{\sc i} \(\lambda1304\),
C\,{\sc ii} \(\lambda1335\), Si\,{\sc iv} \(\lambda1397\),
C\,{\sc iv} \(\lambda1549\), He\,{\sc ii} \(\lambda1640\),
O\,{\sc iii}] \(\lambda1663\), N\,{\sc iii}] \(\lambda1750\),
Si\,{\sc iii}] \(\lambda1883\), C\,{\sc iii}] \(\lambda1908\),
C\,{\sc ii}] \(\lambda2326\), Mg\,{\sc ii} \(\lambda2799\), and
Mg\,{\sc i} \(\lambda2853\). We retained pixels with valid spectral
flags and positive rest-frame wavelengths and flux densities. When an
error spectrum was available, we additionally required finite positive
flux-density uncertainties and \({\rm S/N}\geq1\). Measurements were
accepted only when at least 10 valid continuum pixels remained after the
emission-line masking. The UV continuum was modelled as a power law, $f_\lambda \propto \lambda^{\beta_{\rm UV}}$, fitted directly in linear $f_\lambda$ space. 

We then measured the H${\alpha}$ fluxes and equivalent widths for each of the 37,528 NIRSpec spectra. For each source, the expected observed-frame H${\alpha}$ wavelength was computed from the spectroscopic redshift as $\lambda_{\mathrm{H}\alpha,\mathrm{obs}} = 0.65628(1+z)\,\mu\mathrm{m}$. 
We fitted the spectrum with a linear continuum plus a single Gaussian emission-line profile. The initial line window was set to $\pm0.01\,\mu{\rm m}$ around the expected H$\alpha$ wavelength, while the continuum was estimated from two side bands of width $0.02\,\mu{\rm m}$, separated from the line window by $0.01\,\mu{\rm m}$. When fewer than 10 line pixels or 15 continuum pixels were available, the line and continuum windows, together with their separation, were expanded iteratively by a factor of 1.5, up to six times. A preliminary linear continuum fit was sigma-clipped once (at 3$\sigma$), using a median-absolute-deviation estimate of the residual scatter, before deriving the final continuum model. The H$\alpha$ flux was then obtained from the analytic integral of the best-fitting Gaussian profile. The equivalent width was computed by integrating the continuum-subtracted line profile divided by the fitted continuum over the line window. Finally, the observed-frame EW was converted to rest-frame ($EW_{\rm rest}=EW_{\rm obs}/(1+z)$). 

Finally, for the optical continuum, we modelled the rest-frame continuum over $3646$--$7000\,$\AA\, directly in linear $f_\lambda$ space using an inverse-variance-weighted power law,
$f_\lambda=A(\lambda/5000\,\AA)^{\beta_{\rm opt}}$. We masked the regions around [O\,{\sc ii}] $\lambda3727$, [Ne\,{\sc iii}] $\lambda3869$, H$\delta$, H$\gamma$, H$\beta$, [O\,{\sc iii}] $\lambda\lambda4959,5007$, He\,{\sc i} $\lambda5876$, [O\,{\sc i}] $\lambda6300$, [N\,{\sc ii}] $\lambda\lambda6548,6583$, H$\alpha$, and [S\,{\sc ii}] $\lambda\lambda6716,6731$. In addition to the individual line masks, we excluded the broad H$\beta$+[O\,{\sc iii}] and H$\alpha$+[N\,{\sc ii}]+[S\,{\sc ii}] complexes. We retained valid spectral pixels with finite positive wavelengths, finite fluxes of either sign, and finite positive flux uncertainties. No per-pixel S/N threshold or positive-flux requirement was imposed. Fits were retained when at least 10 continuum pixels remained after masking. The resulting $\beta_{\rm opt}$ measurements therefore trace the rest-frame optical continuum while minimising the contribution from strong nebular emission lines. 

\subsection{Sample selection criteria}
\label{sec:final_selection}
With this in mind, we adopt a two-stage procedure to construct the LRD and the LBD samples. We first identify a parent population of compact, UV blue, strong-H$\alpha$ emitters using criteria that are independent of the rest-frame optical continuum. We then classify these sources as LRDs or LBDs according to their rest-frame optical continuum slope,  $\beta_{\rm opt}$. 

Based on the photometric and spectroscopic measurements described in Section \ref{sec: photometry} and \ref{sec: spectra observations}, respectively, the parent sample is defined using the following criteria:

\begin{enumerate}
        
    \item A rest-frame H$\alpha$ equivalent width in the range 
    \hbox{$200$\,\AA $< \mathrm{EW}_{\mathrm{H}\alpha} < 2000$\,\AA}.
    
    \item A conservative blue UV continuum slope 
    $\beta_{\rm UV} < -0.5$.
    
    \item A compact morphology, defined via the flux ratio 
    $\mathrm{F444W}({0.2^{\prime\prime})}/\mathrm{F444W}({0.1^{\prime\prime})} < 1.7$, consistent with the compactness criteria adopted in previous
    LRD studies \citep[see, e.g.][]{deGraaff2025RUBIES}.
\end{enumerate}

These three criteria yield an initial sample of 555 sources. This parent sample is designed to include both red and blue compact AGN-like sources with prominent H$\alpha$ emission, without imposing any requirement on the shape of the rest-frame optical continuum. The final sample is dominated by low-resolution NIRSpec/PRISM spectroscopy ($R\sim100 - 500$), with only three sources having medium-resolution spectra ($R\sim1000$). This selection is inherently biased towards sources for which both strong H$\alpha$ emission and a measurable UV continuum slope can be identified, and therefore does not represent an unbiased sample of the broader AGN population.

We subsequently use $\beta_{\rm opt}$ to split the sources into LRDs and LBDs (see Section \ref{sec:beta_slope_measurements} for the measurement methodology).
Specifically, we compare their locations in the $\beta_{\rm UV}$--$\beta_{\rm opt}$ plane with the LRD/LBD regions defined by \cite{Brazzini2026}: sources with $\beta_{\rm opt}>0$ are classified as LRD candidates, while those with $\beta_{\rm opt}\leq 0$ are classified as LBD candidates.

We then apply additional quality cuts to ensure robust photometric and optical-slope measurements. We required ${\rm S/N}>5$ for the F444W fluxes measured within both the $0.1^{\prime\prime}$ and $0.2^{\prime\prime}$ apertures (168 sources removed). 
We further required ${\rm F444W \,mag}<27$ for LRD candidates (18 additional sources removed). We also required a reliable $\beta_{\rm opt}$ classification, defined by the expected sign being recovered in at least 95\% of 1000 MC refits: $P(\beta_{\rm opt}>0)\geq0.95$ for all LRDs and $P(\beta_{\rm opt}<0)\geq0.95$ for LBDs (76 sources removed). Finally, we inspected the repeated spectral observations and retained a single measurement per unique source (12 duplicate entries removed), keeping the observation with the higher signal-to-noise ratio.

After applying all quality cuts, the final robust sample comprises 280 sources: 89 LRDs in total, including 72 previously reported LRDs from \citet{deGraaff2025DJA} (68\% recovery fraction), 17 newly identified LRDs, and 191 LBDs. The final selection is drawn from the surveys described in the Appendix Table \ref{tab:survey_summary}. Hence, our sample LRDs comprise $\simeq 32\%$ of the final sample, consistent with the $10-30\%$ reported by \citet{Brazzini2026}.

\begin{figure}
	\includegraphics[width=\columnwidth]{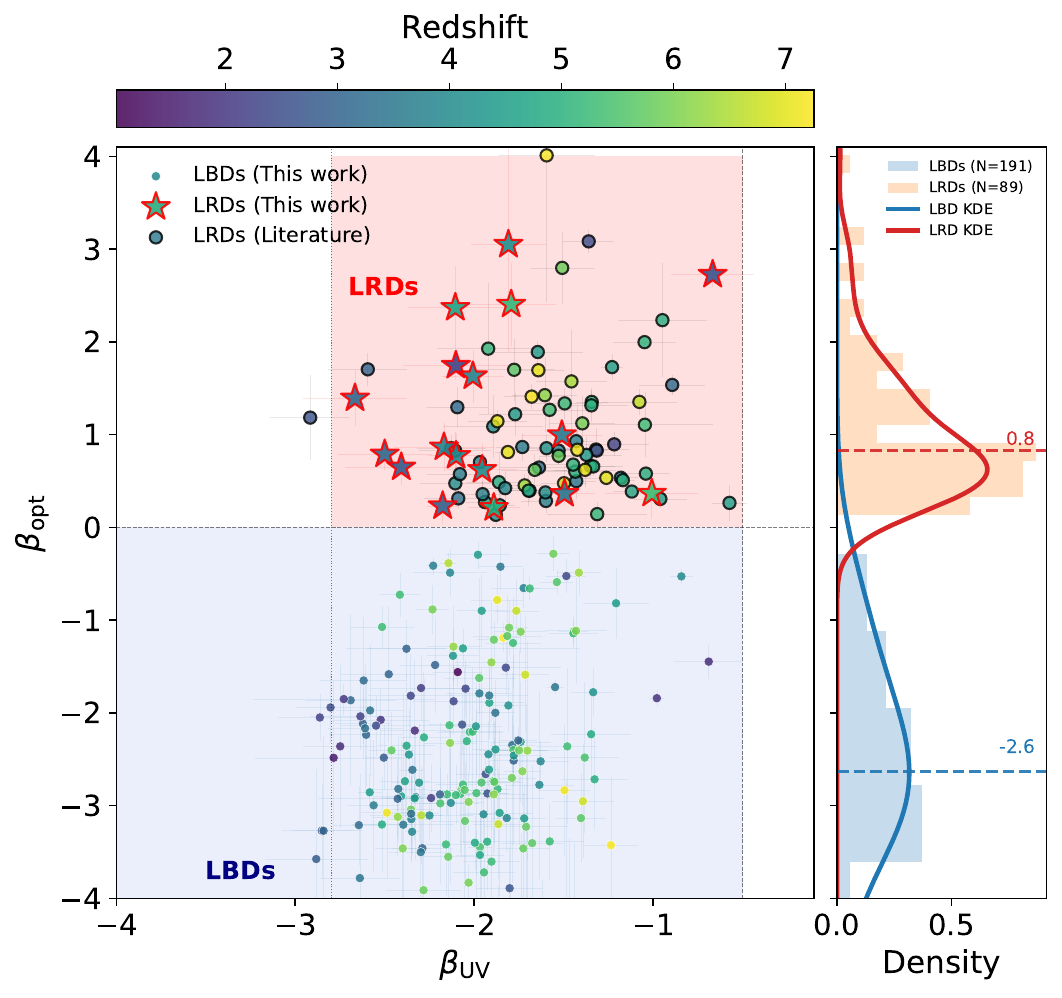}
\caption{Rest-frame optical continuum slope, $\beta_{\rm opt}$, as a function of the UV continuum slope, $\beta_{\rm UV}$, for the selected spectroscopic sample. Points are colour-coded by spectroscopic redshift, as indicated by the colour bar at the top. The red- and blue-shaded regions mark the LRD and LBD classification regions, respectively, following \citet{Brazzini2026}. Red stars indicate the 17 newly identified LRDs from this work, while red-edged circles show the previously reported  LRDs \citep{deGraaff2025DJA}; the remaining blue-continuum sources are classified as LBDs. Error bars indicate the uncertainties on the fitted continuum slopes. 
The right-hand panel shows the corresponding $\beta_{\rm opt}$ distributions and kernel density estimates for the 89 LRDs and 191 LBDs. Dashed horizontal lines mark the median values, $\beta_{\rm opt}=0.8$ for LRDs and $\beta_{\rm opt}=-2.6$ for LBDs.
}
    \label{fig:betaUV_vs_betaOPT}
\end{figure}

Fig.\,\ref{fig:betaUV_vs_betaOPT} shows the resulting classification in the $\beta_{\rm UV}$--$\beta_{\rm opt}$ plane. For all sources, including those previously identified as LRDs in the literature, both continuum slopes were remeasured from the spectra using the homogeneous procedure described in Section~\ref{sec:beta_slope_measurements}. Hence, the distributions are directly comparable. The underpopulated region around $\beta_{\rm opt}=0$ results from our requirement that the slope sign be recovered in at least 95\% of the Monte Carlo realisations. Sources that do not satisfy this criterion, including 20 LRDs from \citet{deGraaff2025DJA}, are excluded from the final sample and may represent a transitional population (see Section \ref{Section: Discussion}). 

We quantify the continuum-slope distributions of the two sub-samples by computing their median values and estimating the associated uncertainties from 10,000 bootstrap resamples. The LRDs have a median optical slope of $\beta_{\rm opt}=0.83^{+0.03}_{-0.05}$, while the LBDs show a markedly bluer distribution, with $\beta_{\rm opt}=-2.63^{+0.15}_{-0.12}$. The median separation between the two populations is therefore $\Delta\beta_{\rm opt}=3.46^{+0.13}_{-0.17}$. As $\beta_{\rm opt}$ enters the classification itself, this offset should not be interpreted as an independent significance test; rather, it quantifies the extent to which the two distributions remain separated from the adopted boundary. A smaller but still significant difference is also present in the UV slopes, with median values of $\beta_{\rm UV}=-1.60^{+0.07}_{-0.06}$ for LRDs and $\beta_{\rm UV}=-1.97^{+0.02}_{-0.05}$ for LBDs.


\subsection{Newly identified LRDs}
\label{Section: newLRDs}

Here, we present the 17 LRDs newly identified through our homogeneous, self-consistent selection, and examine their continuum properties relative to the previously known population.  All candidates were visually inspected and lie within the LRD region of the $\beta_{\rm UV}$--$\beta_{\rm opt}$ plane shown in Fig.\,\ref{fig:betaUV_vs_betaOPT}.

\begin{figure}
\centering
\includegraphics[width=\columnwidth]{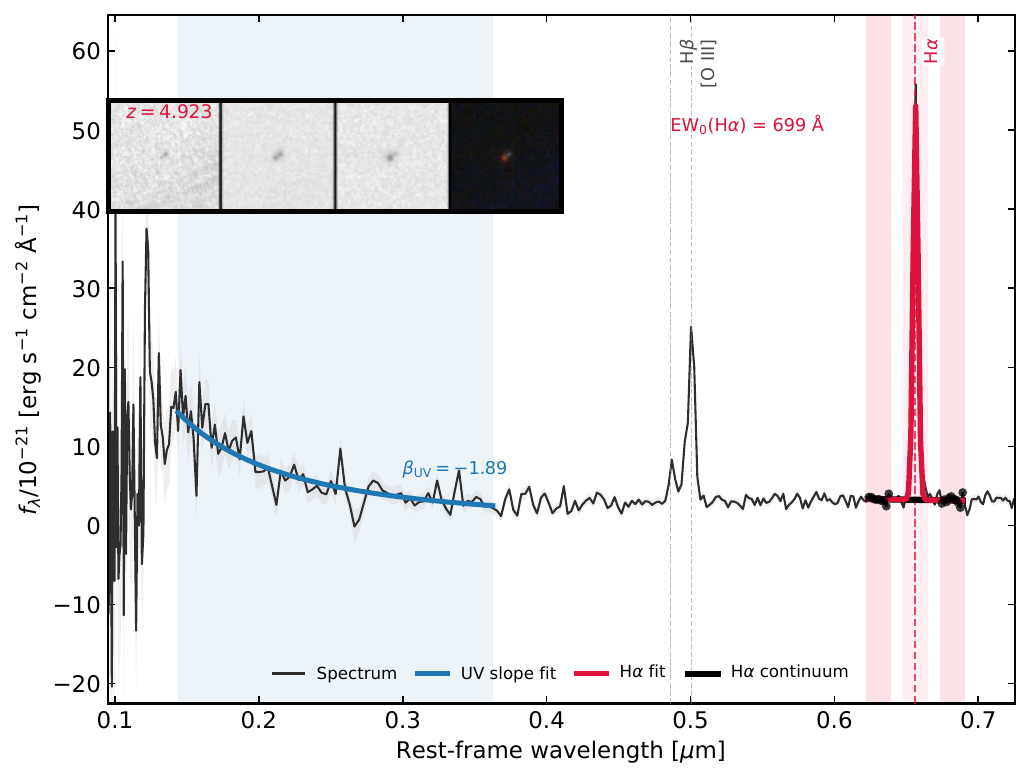}    \includegraphics[width=\columnwidth]{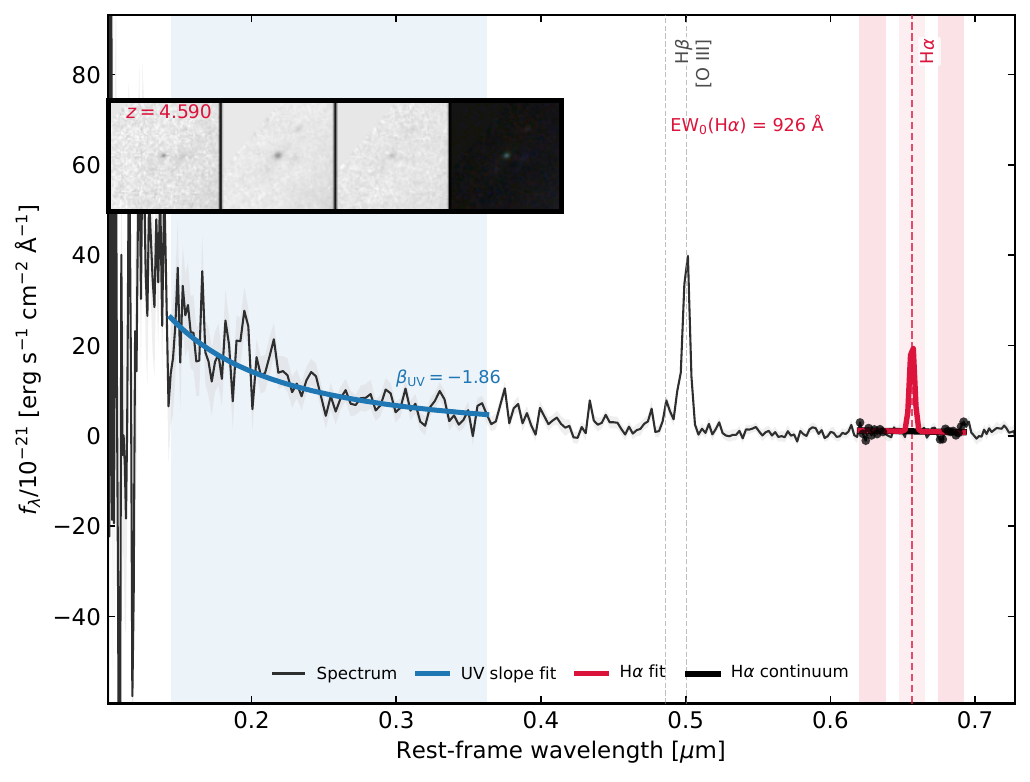}
    \caption{Example spectra of one of the 17 newly-identified LRDs uncovered in this study, shown in the upper panel, and a representative LBD, shown in the lower panel. In each case, the spectrum is shown in black, while the fitted H$\alpha$ emission line and underlying local continuum used to measure the rest-frame H$\alpha$ equivalent width are shown in red. 
The red-shaded regions mark the wavelength intervals used in the H$\alpha$ EW measurement. The UV continuum fit is shown in blue over the short-wavelength region of the spectrum, with the blue shaded region marking the wavelength range used to derive $\beta_{\rm UV}$. The inset panels show the NIRCam F090W, F200W, and F444W imaging (used to assess the compactness criterion and to perform the size measurements), together with the combined RGB colour image.
The newly identified LRD displays a blue UV slope together with a strong H$\alpha$ emission line. At the same time, the representative LBD shows a more prominent blue UV slope, a substantially weaker H$\alpha$ line, and a flatter optical continuum.
}
    \label{fig:newLRD}
\end{figure}

Fig.\,\ref{fig:newLRD} compares a newly identified LRD with a representative LBD selected from the same parent sample. While both sources exhibit blue UV continua, the LRD shows a redder optical continuum and more prominent H$\alpha$ emission. 
The newly identified LRDs generally exhibit less extreme UV-optical slopes than previously reported systems. Their more moderate continuum slopes may help explain their absence from previous LRD selections and support a heterogeneous population spanning a sequence of physical properties \citep{matthee2026_photometricSequence, PerezGonzalez2026}. In Section~\ref{sec: results}, we investigate whether the differences in optical continuum and H$\alpha$ properties between LRDs and LBDs reflect distinct physical regimes or can be attributed to different phases or orientations of the broader AGN population.

\subsection{X-ray counterparts}
\label{sec:xray_crossmatch}

We cross-matched the 280 sources in our final sample with the 5XMM-DR15 \citep{Webb2026} and Chandra Source Catalogue (CSC~2.1; \citealt{Evans2024}), using matching radii of 6 arcsec and 2 arcsec, respectively. We find only two secure X-ray counterparts, both LBDs: a spectroscopically confirmed Seyfert~1 AGN detected in multiple XMM-Newton observations, and a source coincident with a known QSO at $z=3.44$. Other candidate associations are either low-significance detections, unrelated foreground sources, or too faint and/or offset to be considered secure. We therefore find no robust
X-ray detections among the LRDs and two among the LBDs. Given the inhomogeneous X-ray coverage and sensitivity, these results do not allow us to conclude that there are intrinsic differences between the two populations, but are broadly consistent with previous reports of X-ray weakness among ${\it JWST}$-selected AGN \citep{Maiolino2025} and LRDs \citep{Ananna2024}.


\section{Unveiling the Nature of LRDs and LBDs}
\label{sec: results}

The contrasting spectral properties of LRDs and LBDs identified in Section \ref{sec:final_selection} raise the question of whether they represent intrinsically distinct populations, different evolutionary phases, or a single population observed along different lines of sight. In the super-Eddington unification model proposed by  \citet{Madau2026}, LRDs are interpreted as the dust-reddened, high-inclination counterparts of the less-obscured LBDs, with both classes powered by the same underlying AGN engine. By contrast, other scenarios associate the two populations with different phases of galaxy-SMBH co-evolution, potentially driven by differences in obscuration, accretion state, or host-galaxy properties \citep{Asada2026, Brazzini2026}. We test these alternatives through a homogeneous comparison of their continuum, emission-line, and structural properties.

\subsection{H$\alpha$ constraints on a purely orientation-driven LRD-LBD scenario}

We first compare the H$\alpha$ emission properties of LRDs and LBDs, focusing on the line FWHM, luminosity, and rest-frame equivalent width. These quantities provide complementary constraints on the line-emitting region: the FWHM traces the characteristic gas kinematics (albeit potentially with a contribution from scattering), while the luminosity and equivalent width quantify the strength of the line emission, both intrinsically and relative to the underlying continuum. In the simplest orientation-only scenario, in which LRDs and LBDs are drawn from the same intrinsic population and differ primarily in viewing angle, their H$\alpha$ luminosity distributions should be broadly comparable, aside from attenuation or anisotropy effects. Together, these observations allow us to test whether the differences between the two populations can be explained primary by a viewing geometry or instead require differences in their intrinsic emission-line properties. All H$\alpha$ measurements were obtained using the homogeneous fitting procedure described in Section~\ref{sec:beta_slope_measurements}.

\begin{figure}
\centering
	\includegraphics[width=0.9\columnwidth]{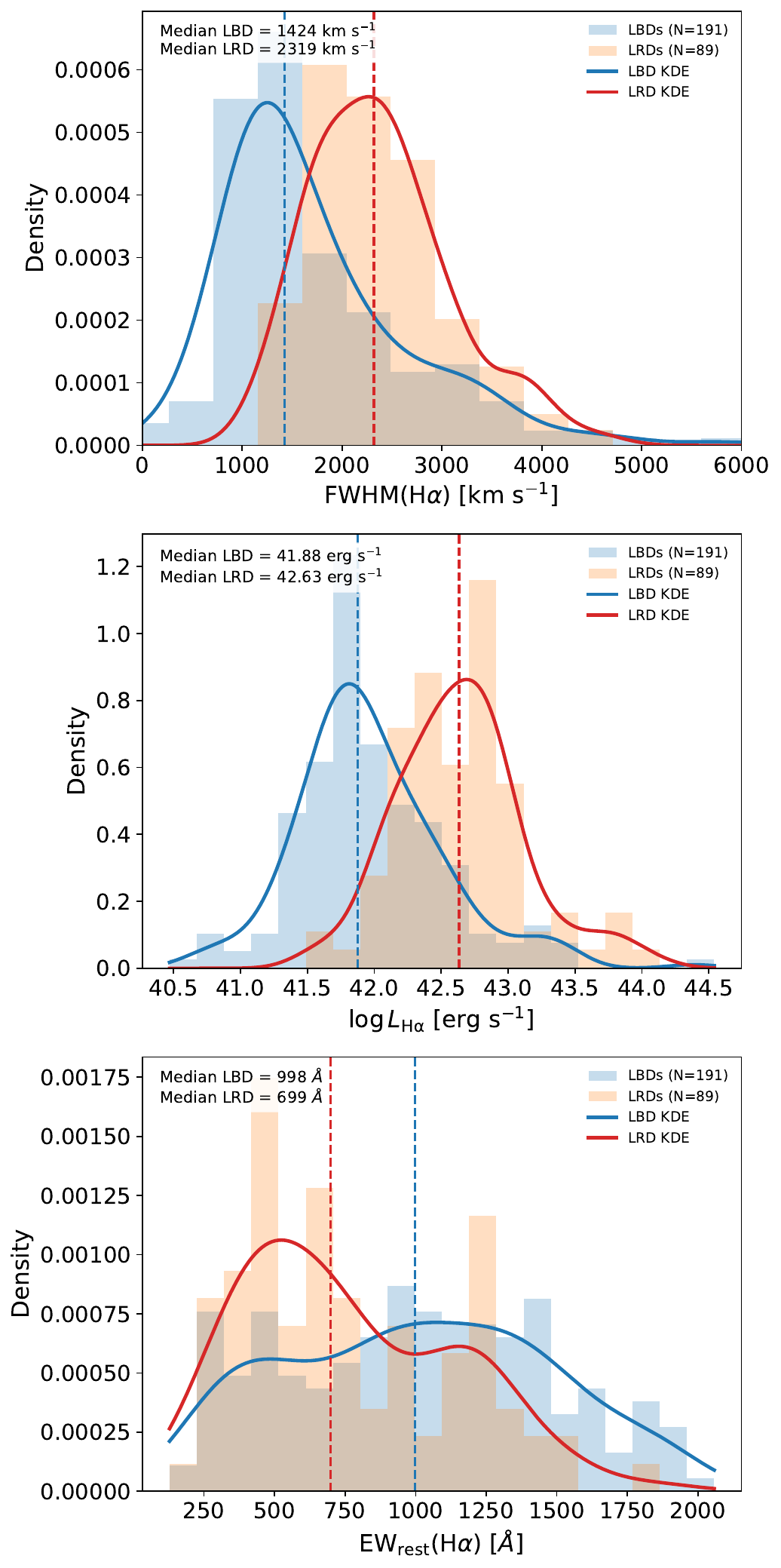}
  \caption{Normalised histograms and kernel density estimates (KDEs) of the H$\alpha$ FWHM, luminosity, and rest-frame equivalent width for the 89 LRDs and 191 LBDs in our sample. LRDs and LBDs are shown in red and blue, respectively, and dashed vertical lines mark the median values. The FWHM distributions overlap, but LRDs are systematically shifted towards broader lines, with median values of $2319^{+71}_{-64}$ and $1424^{+56}_{-43}~{\rm km,s^{-1}}$ for LRDs and LBDs, respectively. An even stronger separation is observed in H$\alpha$ luminosity, with median values of $\log(L_{\rm H\alpha}/{\rm erg,s^{-1}})=42.634^{+0.091}_{-0.092}$ for LRDs and $41.877^{+0.041}_{-0.010}$ for LBDs. The equivalent-width distributions show substantial overlap within the common selection interval, $200<{\rm EW}{\rm rest}({\rm H}\alpha)<2000$~\AA, although LRDs have a lower median EW, $699^{+36}_{-42}$~\AA, than LBDs, $998^{+51}_{-29}$~\AA. Thus, LRDs are characterised by broader and substantially more luminous H$\alpha$ emission, while occupying a broadly comparable EW range imposed by the parent-sample selection.
}
   \label{fig:KDE_FWHM_LHa_EW_LRD_vs_LBD}
\end{figure}

Fig.\,\ref{fig:KDE_FWHM_LHa_EW_LRD_vs_LBD} compares the H$\alpha$ FWHM, luminosity, and rest-frame equivalent-width distributions of the 89 LRDs and 191 LBDs in the final sample. Unless stated otherwise, uncertainties on the median values were estimated by resampling each population with replacement 10,000 times, with the 16th and 84th percentiles of the bootstrap-median distribution defining the lower and upper uncertainties, respectively.

It can be seen that the LRDs are systematically shifted towards broader H$\alpha$ line widths, with median values of $2319^{+71}_{-64}~{\rm km\,s^{-1}}$ and $1424^{+56}_{-43}~{\rm km\,s^{-1}}$ for LRDs and LBDs, respectively. The corresponding median offset is $\Delta{\rm FWHM}=895^{+73}_{-95}~{\rm km\,s^{-1}}$,
and the two distributions are statistically distinct ($p_{\rm KS}=3.42\times10^{-14}$). The H$\alpha$ luminosity distributions show an even stronger separation; LRDs are systematically more luminous in H${\alpha}$ than LBDs, with median values of $\log (L_{\mathrm{H}\alpha}/{\rm erg,s^{-1}})=42.634^{+0.091}_{-0.092}$ and $41.877^{+0.041}_{-0.010}$, respectively. The corresponding offset, $\Delta\log L_{\mathrm{H}\alpha}=0.757^{+0.090}_{-0.099}$ dex, implies that LRDs are, on average, a factor of $\simeq 6$ times 
more luminous in their observed H$\alpha$ line emission. The two distributions are statistically distinct ($p_{\rm KS}=2.01\times10^{-20}$), showing that LRDs effectively occupy the high-luminosity tail of the compact sample. Although the FWHM distributions also differ significantly, the luminosity contrast is the more pronounced distinction between the two populations.

The H$\alpha$ equivalent-width distributions show a much more substantial overlap within the common selection interval, $200<\mathrm{EW}{\rm rest}(\mathrm{H}\alpha)<2000$~\AA. LRDs have a moderately lower median equivalent width ($\mathrm{EW}{(\rm H\alpha)}=699^{+36}_{-42}$~\AA) than LBDs, which have $\mathrm{EW}{\rm (H\alpha)}=998^{+51}_{-29}$~\AA. However, because the H$\alpha$ EW is part of the parent sample selection, this difference should not be interpreted as an independent physical distinction between the two populations. The key result is instead that LRDs are substantially more luminous in H$\alpha$. 

Taken together, these results show that LRDs exhibit broader H$\alpha$ profiles and, more notably, substantially higher H$\alpha$ luminosities than LBDs. This luminosity offset is unlikely to be driven by differences in redshift coverage. The two samples span largely overlapping redshift ranges, with median values of $z_{\rm med}=4.53^{+0.30}_{-0.30}$ for LRDs and $z{\rm med}=4.80^{+0.27}_{-0.21}$ for LBDs. Although the redshift distributions differ modestly, 
all sources have reliable spectroscopic redshifts, and the LRDs in our sample do not preferentially lie at higher redshift. The enhanced H$\alpha$ luminosity of LRDs therefore cannot be attributed solely to a systematic redshift difference between the two samples. 

Given the heterogeneous definitions of LBDs adopted in the literature, we repeated the comparison after imposing the same broad-line criterion, ${\rm FWHM}({\rm H}\alpha)>2000~{\rm km,s^{-1}}$, on both populations. The H$\alpha$ luminosity offset persists and increases, reaching $\Delta\log L_{\rm H\alpha} \sim 1$ dex 
corresponding to a factor of $\sim10$. 
Thus, the enhanced H$\alpha$ luminosity of LRDs is not driven by differences in redshift coverage or by the adoption of different broad-line thresholds. We discuss the implications of this result, and the impact of the heterogeneous LRD and LBD definitions used in the literature, in Section~\ref{Section: Discussion}.


In the simplest orientation-only scenario, the probability of observing a source as an LRD should be approximately independent of the intrinsic luminosity. Since broad H$\alpha$ luminosity provides an observational proxy for bolometric luminosity in broad-line AGN, we therefore test explicitly whether the LRD fraction varies systematically with $L_{\rm H\alpha}$. 

\begin{figure}
	\includegraphics[width=\columnwidth]{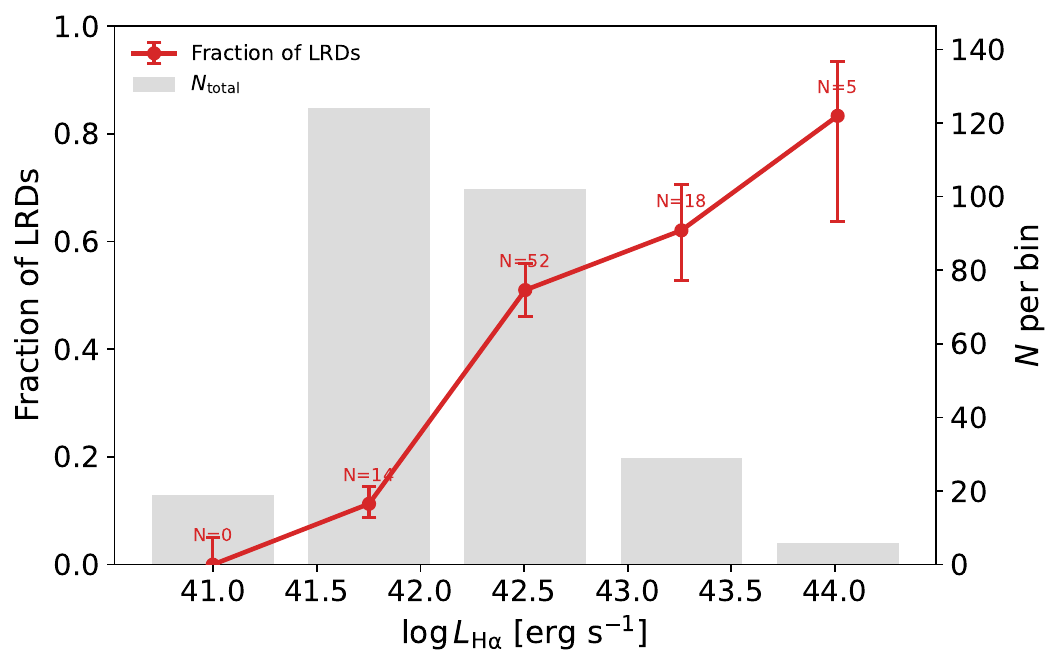}
\caption{LRD fraction as a function of H$\alpha$ luminosity for the final sample of 89 LRDs and 191 LBDs. Red points show the fraction $f_{\rm LRD}=N_{\rm LRD}/(N_{\rm LRD}+N_{\rm LBD})$ in independent bins of $\log L_{\rm H\alpha}$, with error bars indicating binomial confidence intervals. Grey bars show the total number of sources in each bin, referenced to the right-hand axis, while the labels indicate the corresponding number of LRDs. The LRD fraction increases strongly with H$\alpha$ luminosity, from $f_{\rm LRD}=0.11$ at $\log L_{\rm H\alpha}=41.75$ to $0.51$ at $\log L_{\rm H\alpha}=42.51$ and $0.83$ in the highest-luminosity bin. 
The overall trend shows that LRDs become increasingly prevalent towards the high-$L_{\rm H\alpha}$ tail of the compact-source population. }

    \label{fig:LRD_fraction_vs_logLHa}
\end{figure}

Fig.\,\ref{fig:LRD_fraction_vs_logLHa} shows the LRD fraction in five independent luminosity bins centred at $\log L_{\rm H\alpha}=41.00$, 41.75, 42.51, 43.26, and 44.01. The uncertainties are computed using binomial confidence intervals. The LRD fraction increases strongly with H$\alpha$ luminosity. No LRDs are found in the lowest-luminosity bin ($0/19$), while LRDs account for $14/124$ sources ($f_{\rm LRD}=0.11$) at $\log L_{\rm H\alpha}=41.75$. The fraction rises sharply to $52/102$ ($f_{\rm LRD}=0.51$) at $\log L_{\rm H\alpha}=42.51$ and to $18/29$ ($f_{\rm LRD}=0.62$) at $\log L_{\rm H\alpha}=43.26$. In the highest-luminosity bin, five of the six sources are LRDs, corresponding to $f_{\rm LRD}=0.83$, although this value is subject to substantial small-number uncertainty. Thus, LRDs transition from being rare at low H$\alpha$ luminosities to constituting the majority of the population above $\log L_{\rm H\alpha}\simeq42.5$, indicating that they preferentially occupy the high H$\alpha$-luminosity tail of the compact-source population.

This trend is hard to explain in the simplest luminosity-independent orientation scenario. If LRDs and LBDs are drawn from the same intrinsic population and differ only in viewing angle, the fraction observed as LRDswould not be expected to increase strongly or even reduce somewhat with increasing observed H$\alpha$ luminosity (due to increasing attenuation). Using $L_{\rm H\alpha}$ as an empirical proxy, the observed increase in $f_{\rm LRD}$ therefore suggests that the LRD phenomenon is preferentially associated with systems which produce intrinsically more luminous H$\alpha$ emission. This may reflect intrinsic differences in accretion state, obscuration geometry, central engine properties, or host galaxy conditions, although attenuation, anisotropy, and selection effects may also influence the observed H$\alpha$ luminosity.

\begin{figure*}
    \centering
    \includegraphics[width=1\linewidth]{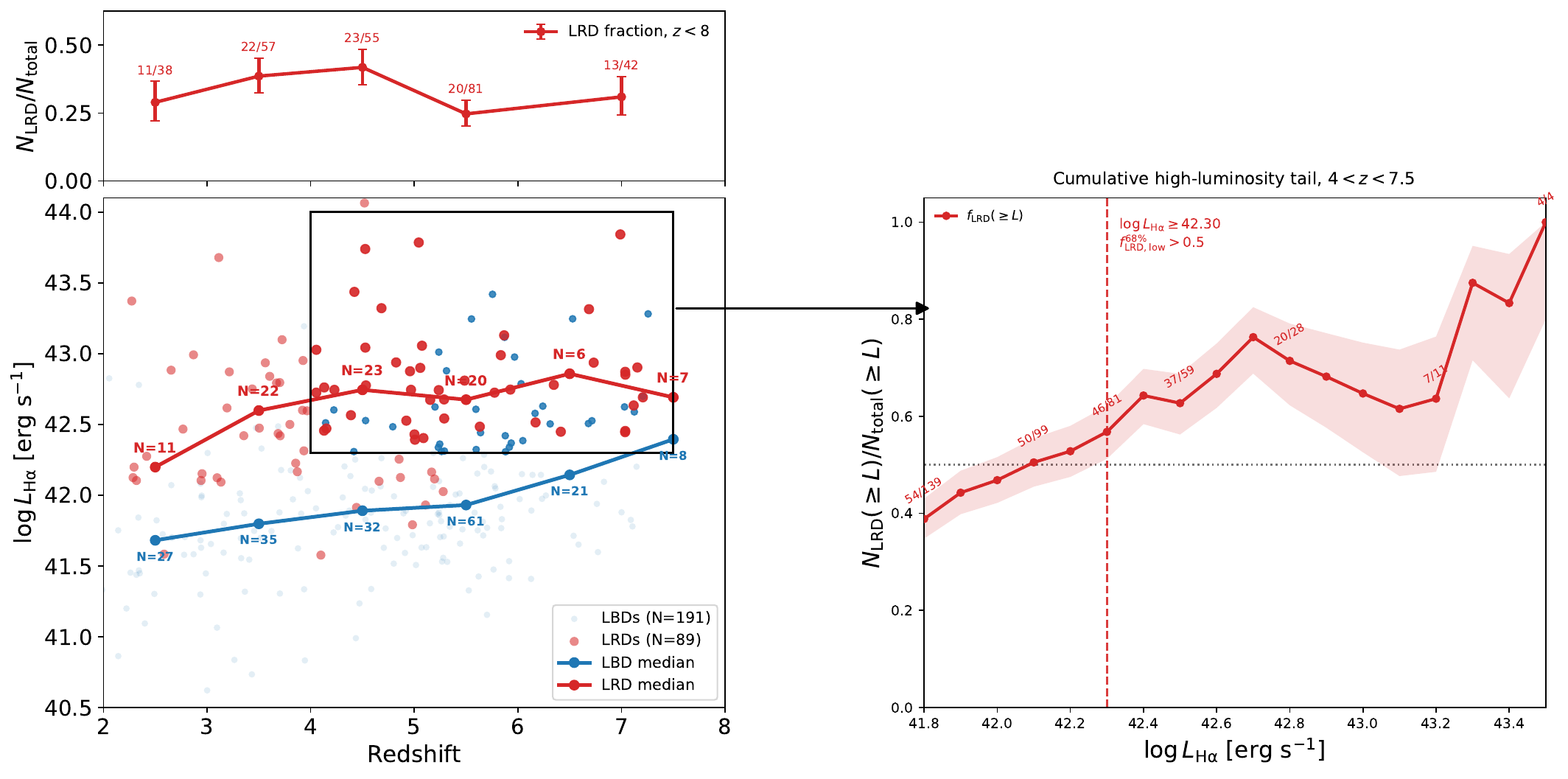}
    \caption{
    H$\alpha$ luminosity as a function of redshift for LRDs and LBDs.
    Individual sources are shown as red and blue points, respectively, while the red and blue lines indicate the median $\log L_{\rm H\alpha}$ in each redshift bin.
    The number of sources contributing to each median bin is labelled in the lower-left panel.
    The upper-left panel shows the fraction of LRDs relative to the total LRD+LBD population as a function of redshift, computed for $z<8$.
    The LRD fraction shows no clear monotonic evolution with redshift, whereas LRDs remain systematically more H$\alpha$ luminous than LBDs throughout the redshift range considered.
    The rectangle highlights the high-redshift, high-luminosity region connected to the cumulative analysis shown in the right-hand panel.
    The right-hand panel shows the cumulative LRD fraction above a given luminosity threshold, $f_{\rm LRD}(\geq L)=N_{\rm LRD}(\geq L)/N_{\rm total}(\geq L)$, computed within $4<z<7.5$.
    The shaded region indicates the 68\% Wilson binomial confidence interval.
    The cumulative LRD fraction generally increases towards higher luminosity thresholds, showing that LRDs become progressively over-represented in the high-$L_{\rm H\alpha}$ tail.
    The vertical dashed line at $\log L_{\rm H\alpha}=42.30$ marks the first threshold at which the lower bound of the 68\% Wilson confidence interval exceeds 0.5, indicating that LRDs constitute a statistically robust majority above this luminosity.
    The highest-luminosity measurements are based on only a few sources and therefore have comparatively large uncertainties.
    }
    \label{fig:LHalpha_vs_redshift_cum}
\end{figure*}

To investigate whether LRDs and LBDs occupy different evolutionary phases, we compare their H$\alpha$ luminosities as a function of redshift in Fig.~\ref{fig:LHalpha_vs_redshift_cum}. The sample used in this analysis is again our final sample, which contains 89 LRDs and 191 LBDs with reliable H$\alpha$ luminosities. At fixed redshift, LRDs are systematically more H$\alpha$ luminous than LBDs, with the median offset remaining visible across the full redshift range of $2<z<8$. The LRD fraction, however, shows no clear monotonic evolution with redshift. It increases from $11/38$ ($f_{\rm LRD}=0.29$) at $2<z<3$ to $22/57$ ($f_{\rm LRD}=0.39$) at $3<z<4$ and reaches $23/55$ ($f_{\rm LRD}=0.42$) at $4<z<5$. It then decreases to $20/81$ ($f_{\rm LRD}=0.25$) at $5<z<6$, before rising again to $13/42$ ($f_{\rm LRD}=0.31$) over $6<z<8$. Thus, the persistent luminosity offset between LRDs and LBDs does not appear to be driven simply by a monotonic change in their relative abundance with redshift. In the same figure, we also examine how the incidence of LRDs changes along the high-luminosity tail by computing the cumulative LRD fraction above successive H$\alpha$ luminosity thresholds within the redshift range $4<z<7.5$. At $\log L_{\rm H\alpha}\geq41.8$, LRDs account for $54/139$ sources, corresponding to $f_{\rm LRD}=0.39$. This fraction rises as progressively fainter sources are excluded, reaching $50/99$ ($f_{\rm LRD}=0.51$) and $46/81$ ($f_{\rm LRD}=0.57$) at higher luminosity thresholds. The lower bound of the 68\% Wilson confidence interval first exceeds 0.5 at $\log L_{\rm H\alpha}=42.30$, indicating that LRDs constitute a statistically robust majority above this luminosity. The cumulative fraction continues to increase towards the most luminous end, reaching $20/28$ ($f_{\rm LRD}=0.71$), although the highest-luminosity measurements are increasingly affected by small-number statistics, with only $7/11$ and ultimately $4/4$ sources remaining. This complementary cumulative analysis therefore shows that, within the same redshift interval, LRDs become progressively more prevalent towards the high-$L_{\rm H\alpha}$ tail of the compact-source population.

Taken together, the H$\alpha$ results show that the differences between LRDs and LBDs are not limited to their continuum colours. LRDs exhibit broader line profiles and, most notably, substantially higher H$\alpha$ luminosities, and consequently their incidence increases strongly towards the high-luminosity tail of the line luminosity distribution. These trends persist after accounting for the overlapping redshift distributions and after imposing the same broad-line criterion on both populations. They therefore place tension on the simplest luminosity-independent orientation-only scenario, in which the probability of observing a source as an LRD would not be expected to increase with observed H$\alpha$ luminosity. However, attenuation, anisotropic emission, and selection effects may still contribute to the observed differences. The H$\alpha$ results alone therefore do not establish whether LRDs represent a distinct population or a different evolutionary phase, but they indicate that viewing angle alone is unlikely to provide a complete explanation.

\subsection{Enhanced Balmer decrements in LRDs} 

The previous H$\alpha$ analysis indicates that one of the main differences between LRDs and LBDs lies in the strength of H$\alpha$ emission rather than in the characteristic velocity scale of the broad-line gas. This may result from continuum suppression, intrinsic differences in the line-emitting gas, or a combination of these effects. To distinguish between these possibilities, we examine the Balmer decrement of both LRDs and LBDs and compare their strengths with those displayed by low-redshift broad-line QSOs. 

\begin{figure}
	
\includegraphics[width=\columnwidth]{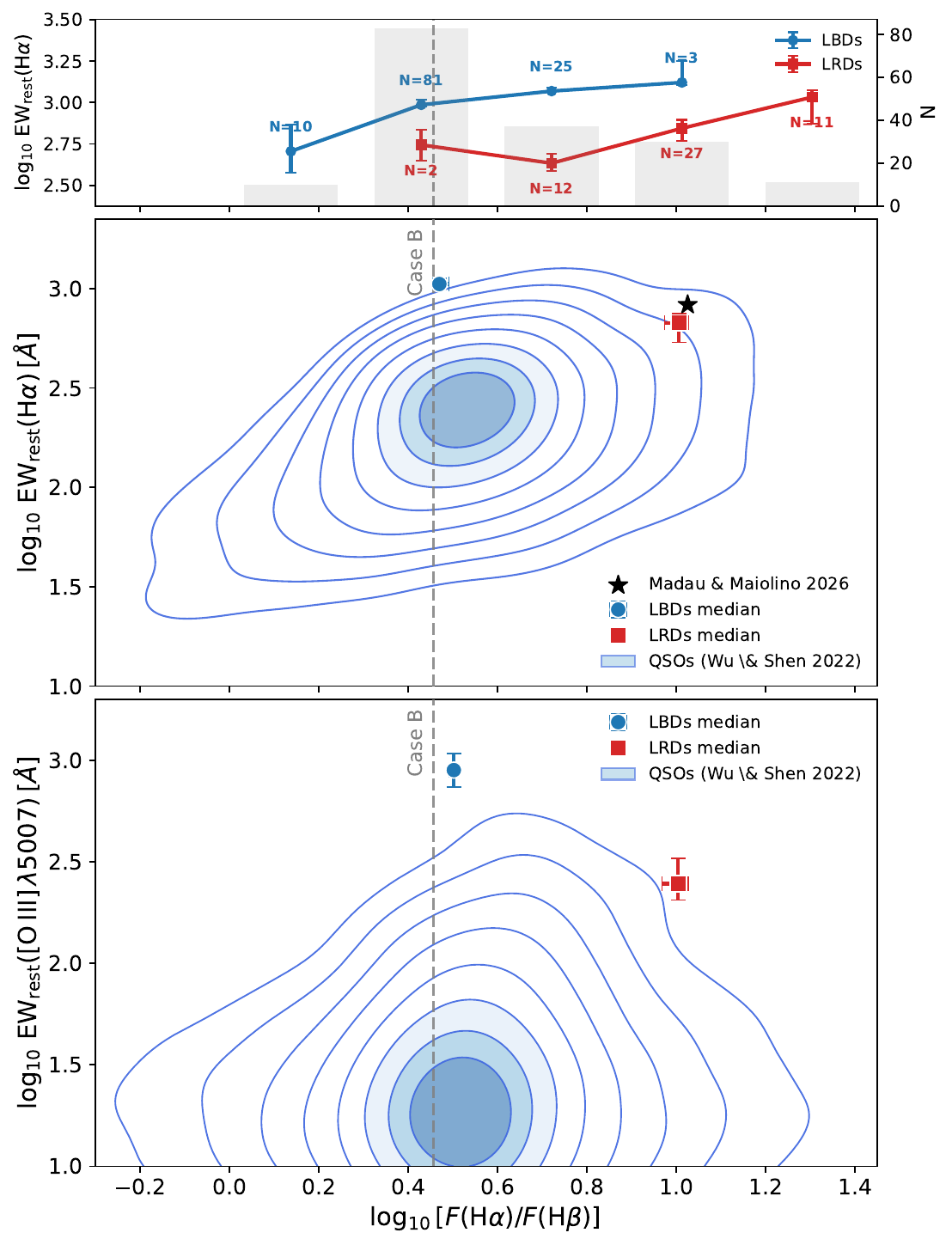} 
 		\caption{Balmer decrement and emission-line equivalent widths for the LBD and LRD populations. 
In all panels, blue circles and red squares show LBDs and LRDs, respectively, with error bars indicating the 16th-84th percentile uncertainties from 1000 bootstrap resamplings of the median. 
The vertical dashed line marks the Case~B recombination value, $F_{\mathrm{H}\alpha}/F_{\mathrm{H}\beta}=2.86$ ($\log_{10}=0.46$), while the contours and shaded regions show the distribution of low-redshift broad-line QSOs from \citet{Wu2022}.
(Top) Median rest-frame H$\alpha$ equivalent width in bins of $\log_{10}(F_{\mathrm{H}\alpha}/F_{\mathrm{H}\beta})$ for the 122 LBDs and 52 LRDs with positive H$\beta$ flux measurements. The labels indicate the number of sources in each bin, and the grey bars show the total number of sources.
(Middle) Median rest-frame H$\alpha$ equivalent width as a function of the Balmer decrement. LBDs have median $F_{\mathrm{H}\alpha}/F_{\mathrm{H}\beta}=2.95$ and ${\rm EW}_{\rm rest}({\rm H}\alpha)=1055$~\AA, compared with $10.14$ and $673$~\AA\ for LRDs, respectively. The black star marks the fiducial model of \citet{Madau2026}, with $F_{\mathrm{H}\alpha}/F_{\mathrm{H}\beta}=10.6$ and ${\rm EW}_{\rm rest}({\rm H}\alpha)=830$~\AA.
(Bottom) Median rest-frame [O\,III]~$\lambda5007$ equivalent width as a function of the Balmer lines, comprising 77 LBDs and 41 LRDs. LBDs have a median ${\rm EW}_{\rm rest}([{\rm O\,III}])=893$~\AA, while LRDs have a substantially lower median of $246$~\AA.}
\label{fig:LRD_LBD_BalmerDecrement_EWHa_binned_QSO_locus}
\end{figure}

Fig.\,\ref{fig:LRD_LBD_BalmerDecrement_EWHa_binned_QSO_locus} shows the distribution of LRDs and LBDs in the plane of Balmer decrement, $\log_{10}(F_{\mathrm{H}\alpha}/F_{\mathrm{H}\beta})$, and rest-frame H$\alpha$ equivalent width. The analysis includes a reduced sample of 174 sources with valid positive H$\alpha$ and H$\beta$ flux measurements, comprising 52 LRDs and 122 LBDs. LRDs have a substantially larger median Balmer decrement than LBDs with $\log_{10}(F_{\mathrm{H}\alpha}/F_{\mathrm{H}\beta})=1.01\pm0.03$ and $0.47\pm0.01$, respectively. In contrast, their median H$\alpha$ EW is lower ($699^{+36}_{-42}$ versus $998^{+51}_{-29}$~\AA). Thus, LRDs preferentially occupy the high-Balmer-decrement region, but not the high-EW tail of the distribution. Within the range covered by the bins, containing all 52 LRDs and 119 of the 122 LBDs, the LRD fraction increases strongly with Balmer decrement. It rises from $0/10$ and $2/83$ ($f_{\rm LRD}=0.02$) in the two lowest populated bins to $12/37$ ($0.32$), $27/30$ ($0.90$), and $11/11$ ($1.00$) in the three highest bins. Although the final bin is sensitive to small-number statistics, the progressive increase demonstrates that LRDs dominate the high Balmer decrement tail of the compact broad-line population.

Both populations are displaced from the central locus of the low-redshift broad-line QSOs of \citet{Wu2022}, indicating that their Balmer-line and continuum properties are not typical of local QSOs. LRDs nevertheless remain within the outer QSO contours, overlapping the high-decrement tail of the local population, whereas LBDs lie close to the contour boundary because of their comparatively large H$\alpha$ EWs. The LRDs may therefore represent more strongly attenuated analogues of the extreme local-QSO population, while the LBD locus may reflect enhanced line emission, a weaker optical continuum, or different line-emitting conditions. The strong-H$\alpha$ selection should, however, be considered when interpreting the large LBD EWs.

The systematically different Balmer decrements also argue against LBDs being simply LRDs with a larger host-galaxy continuum contribution. Such dilution can alter the measured EWs but does not naturally produce the observed difference in $F_{\mathrm{H}\alpha}/F_{\mathrm{H}\beta}$. The systematically different Balmer decrements therefore suggest differences in the attenuation or radiative-transfer conditions affecting their Balmer-emitting gas.  

The median LRD location is consistent with that obtained by the super-Eddington unification model of \citet{Sun2026, Madau2026}. The observed LRD ratio, $F_{\mathrm{H}\alpha}/F_{\mathrm{H}\beta}\simeq10.2$, is close to the model value of 10.6, while its median EW is broadly comparable to the predicted $\simeq830$~\AA. However, the relative EW trend is opposite to that expected from inclination-driven continuum suppression: LRDs have lower, rather than higher, H$\alpha$ EWs than LBDs. Thus, although orientation and dust attenuation may explain the large Balmer decrements of LRDs, they cannot alone reproduce the relative locations of the two populations. Additional differences in their continua, line-emitting regions, or accretion states are therefore required.

\section{Discussion.} 
\label{Section: Discussion}

Our analysis reveals both continuity and systematic differences between LRDs and LBDs. Their continuum and emission-line properties overlap, making individual sources near the classification boundary difficult to assign robustly to either population. Nevertheless, LRDs exhibit broader and substantially more luminous H$\alpha$ emission, larger Balmer decrements, and lower median H$\alpha$ and [O\,III] equivalent widths than LBDs. In this section, we first examine the continuity between the two populations and the limitations of their spectroscopic classification, before considering the implications of their emission-line properties for orientation-based and gas-cocoon models.

The transition between LRDs and LBDs forms a continuous sequence in the $\beta_{\rm UV}$--$\beta_{\rm opt}$ plane, similar to the results of \citet{Asada2026}. This extends the continuous distribution of optical colours and SED properties reported by \citet{Rinaldi2026} and is consistent with the gradual variation in continuum and emission-line properties with optical colour found across LRD subtypes \citep{Barro2026b,PerezGonzalez2026}. We stress that the apparent gap around $\beta_{\rm opt}=0$ in Fig.\,\ref{fig:betaUV_vs_betaOPT} is produced by requiring the sign of $\beta_{\rm opt}$ to be recovered in at least 95 per cent of the Monte Carlo realisations. This criterion preferentially removes sources near the boundary, making transitional objects difficult to classify robustly as either LRDs or LBDs.

The gap itself is therefore artificial, but the distribution before applying this criterion remains bimodal (see Fig. \ref{fig:bimodality_before_betaoptcleaning} in the Appendix). A two-component Gaussian-mixture model is strongly preferred over a single Gaussian, with $\Delta{\rm BIC}=28.65$ and components centred at $\beta_{\rm opt}=-2.53$ and $0.44$. The underlying distribution is better described by two overlapping components rather than either a single unimodal population or two classes separated by a sharp physical boundary. The overlap in their emission-line properties similarly suggests that sources with intermediate optical slopes may represent a genuine transition between the characteristic properties of LBDs and LRDs, although the present data cannot establish whether this corresponds to an evolutionary sequence.

We also investigated whether LRDs can be identified using the emission-line diagnostics developed for faint, narrow-line AGN by \citet{Treiber2025}. As some of the published DJA measurements, particularly the H$\alpha$ fluxes and equivalent widths of LRDs, proved unreliable, we independently refitted the relevant lines and recalculated these quantities for consistency. Even with the revised measurements, our LRDs do not occupy the same line-ratio region of the strongest AGN candidates identified by \citet{Treiber2025}. Weak He\,II emission may instead indicate a softer ionising spectrum rather than the absence of an AGN. Similarly, the two LRDs analysed by \citet{Papovich2026} occupy UV diagnostic regions that are also compatible with star formation or composite ionisation, although their extreme UV equivalent widths and [O\,III]~$\lambda4363$/H$\gamma$ ratios indicate an accreting contribution. Low-resolution line ratios therefore cannot robustly identify individual LRDs as narrow-line AGN. A resolved broad H$\alpha$ component without corresponding broad forbidden-line emission provides stronger evidence for accretion, although the interpretation may remain ambiguous for the faintest broad components \citep{Maiolino2025}.

To make a direct comparison with the recent study of LBDs and LRDs presented by \citet{Geris2026}, it is therefore important to carefully account for differences in sample construction and line measurements. Their study analyses 19 LRDs and 17 LBDs at $2.26<z<7.89$, drawn primarily from the GOODS-N and GOODS-S fields, whereas our sample comprises 89 LRDs and 191 LBDs distributed across multiple extragalactic fields. \citet{Geris2026} also adopt different compactness criteria and spectroscopically classify LBDs as broad-line AGN only when the broad permitted emission has no corresponding broad [O\,III] component, thereby reducing the possibility of an outflow origin \citep[see also][]{Brazzini2026}. We do not impose this requirement because type-I AGN can host ionised outflows, 
and our analysis is not restricted to sources with an unambiguous type-I classification. The two selections may therefore probe somewhat different spectroscopic populations. 

Cross-matching the two catalogues, we recover 14 of the 36 \citet{Geris2026} sources. The remaining objects are excluded mainly because H$\alpha$ is not covered by the available spectra or because they do not satisfy our common selection criterion,
$200~\text{\AA}<EW(\mathrm{H}\alpha)<2000~\text{\AA}$. We find no evidence that the differences between the two samples are driven by field-to-field variations, at least from the comparison of our GOODS-N and GOODS-S subsamples.

An additional methodological difference concerns the definition of $EW(\mathrm{H}\alpha)$. \citet{Geris2026} report the equivalent width of the decomposed broad component, whereas our single-component fits measure the total H$\alpha$ emission. Our cross-matched comparison indicates that this difference has a substantially larger effect for LBDs than for LRDs: the LRD measurements are closer, although still not fully consistent, whereas the LBD equivalent widths are strongly separated. This likely explains much of the discrepancy between the LBD loci obtained in the two studies. Although separating the broad and narrow Balmer components would provide a more physically consistent comparison, most of our spectra were obtained with NIRSpec/PRISM. At this resolution, such a decomposition is strongly model-dependent and cannot be performed robustly and homogeneously across the full sample. We therefore favour a consistent measurement of the total line properties.

Despite these differences, the narrow-line diagnostics of \citet{Geris2026} place LRDs and LBDs in largely overlapping regions consistent with AGN excitation. For $EW([\mathrm{O\,III}])$, they find similar distributions for the two populations, whereas the right-hand panels of Fig.\,\ref{fig:LRD_LBD_BalmerDecrement_EWHa_binned_QSO_locus} show a lower median $EW([\mathrm{O\,III}])$ for our LRDs. Neither result supports the simplest prediction that LRDs should exhibit enhanced $EW([\mathrm{O\,III}])$ if approximately isotropic narrow-line emission is measured against an orientation-dependent continuum as proposed by \citet{Madau2026}. 

\begin{figure}
    \includegraphics[width=\columnwidth]{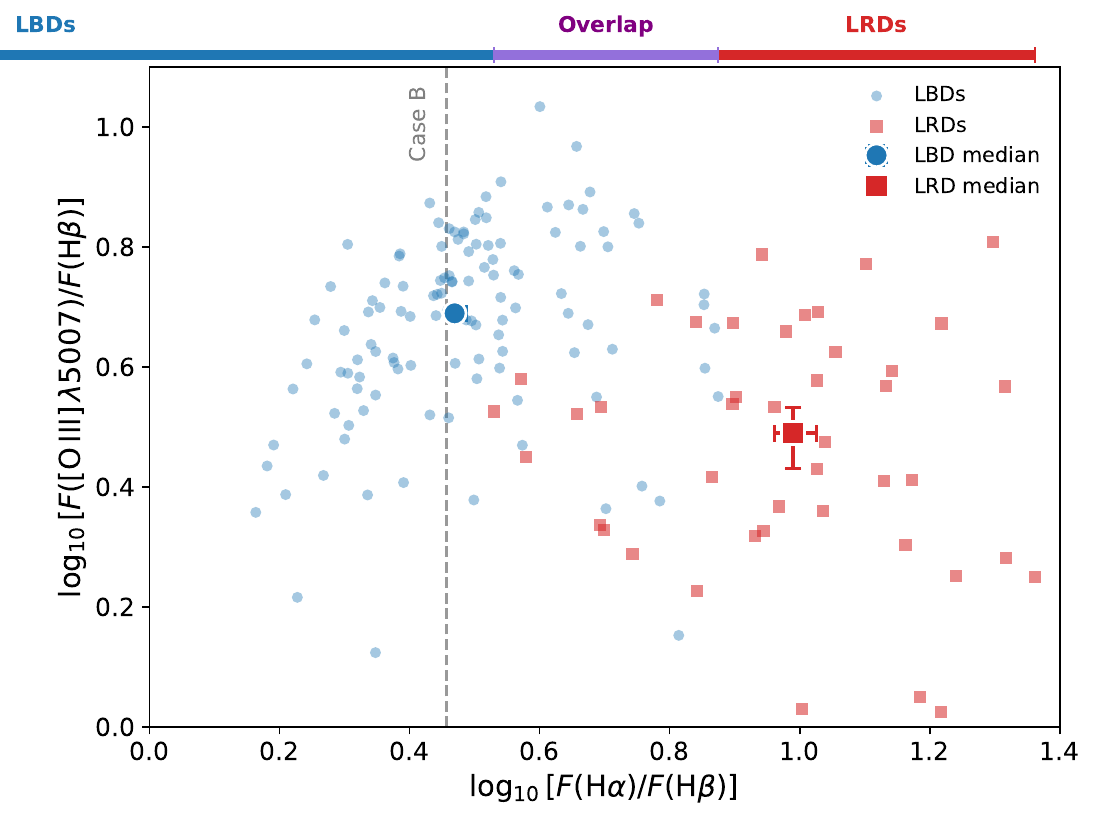}
    \caption{
    [O\,III]~$\lambda5007$/H$\beta$ ratio as a function of the Balmer decrement for LBDs (blue circles) and LRDs (red squares). Large symbols indicate the population medians, with uncertainties estimated through bootstrap resampling, while the vertical dashed line marks the Case~B value, $F_{\mathrm{H}\alpha}/F_{\mathrm{H}\beta}=2.86$. The coloured bars above the panel show the full Balmer-decrement ranges spanned by each population; the region labelled ``Overlap'' denotes their observed overlap in the Balmer decrement. 
    LRDs exhibit systematically larger Balmer decrements but a lower median [O\,III]~$\lambda5007$/H$\beta$ ratio than LBDs. LBDs show a moderate positive Spearman correlation between the two ratios ($\rho=0.38$, $p=1.4\times10^{-5}$), whereas no significant correlation is detected for LRDs ($\rho=-0.06$, $p=0.68$). The LBD correlation should be interpreted cautiously because H$\beta$ enters both axes. 
    Because [O\,III]~$\lambda5007$ and H$\beta$ are closely spaced in wavelength, their ratio is only weakly affected by dust attenuation. Thus, the separation between the population medians is unlikely to be produced by foreground reddening alone, and may instead reflect differences in ionisation conditions, gas density, metallicity, or the visibility of the narrow-line region. This disfavours a simple foreground-screen interpretation of the LRD–LBD distinction and suggests that differences in ionisation conditions or narrow-line-region properties may also play a role.
    }
    \label{fig:OIII5007_Hbeta_vs_BalmerDecrement}
\end{figure}

To determine whether our $EW([\mathrm{O\,III}])$ result is driven by differences in continuum strength, Fig.\,\ref{fig:OIII5007_Hbeta_vs_BalmerDecrement} compares [O\,III]~$\lambda5007$ directly with H$\beta$. Since the two lines are close in wavelength, their flux ratio is largely insensitive to continuum differences and differential attenuation. LRDs have a lower median [O\,III]~$\lambda5007$/H$\beta$ ratio than LBDs, demonstrating that their weaker $EW([\mathrm{O\,III}])$ cannot be attributed solely to their continua. Instead, it indicates differences in the ionisation conditions, gas density, metallicity, or visibility of the narrow-line-emitting region. Nevertheless, the overlap between the two populations mirrors the continuity observed in the $\beta_{\rm UV}$--$\beta_{\rm opt}$ plane, rather than revealing two sharply separated emission-line populations.

LRDs have a median Balmer decrement of $F_{\mathrm{H}\alpha}/F_{\mathrm{H}\beta}=10.1\pm0.7$, substantially larger than the value measured for LBDs. This is consistent with the value of $\simeq8.7$ reported by \citet{deGraaff2025DJA} and the fiducial value of $10.6$ predicted by \citet{Madau2026}, although it is lower than the values of $\sim15$ and $16.2\pm2.2$ measured from the stacked spectra of \citet{Geris2026} and \citet{Sun2026}, respectively. These measurements are not strictly equivalent: \citet{deGraaff2025DJA} measure the unresolved broad-plus-narrow emission; \citet{Geris2026} use the decomposed broad components; and \citet{Sun2026} obtain their value from a host-subtracted stack. Nevertheless, all consistently indicate that large Balmer decrements are a characteristic property of LRDs.

Interpreting these decrements solely as foreground dust attenuation remains challenging given the weak thermal dust emission reported for LRDs \citep{Akins2025,Casey2024} and the possible importance of geometry and radiative-transfer effects. Moreover, because [O\,III] and H$\beta$ experience nearly identical dust attenuation, foreground reddening cannot explain the lower [O\,III]/H$\beta$ ratio of LRDs. Their large Balmer decrements must therefore be considered together with changes in the physical conditions or visibility of the line-emitting gas rather than interpreted simply as a direct measure of a uniform dust screen.

The definition of $EW(\mathrm{H}\alpha)$ must also be considered when comparing our results with the broad-component predictions of \citet{Madau2026}. Nevertheless, the absolute location of our LRDs in the $EW(\mathrm{H}\alpha)$-Balmer-decrement plane is broadly consistent with their fiducial model, as shown in Fig.\,\ref{fig:LRD_LBD_BalmerDecrement_EWHa_binned_QSO_locus}. Our median $EW(\mathrm{H}\alpha)=699^{+36}_{-42}$~\AA\ is $131$~\AA\ ($16\%$) lower than their predicted value of $830$~\AA, while our median Balmer decrement agrees with their fiducial value of $10.6$. Our equivalent widths also broadly agree with the unresolved measurements of \citet{deGraaff2025DJA} and \citet{Sun2026}, who report $762^{+43}_{-60}$~\AA\ and $817.2^{+80.9}_{-82.7}$~\AA, respectively. These values are only $63$~\AA\ ($8\%$) and $118$~\AA\ ($14\%$) above our median, and provide more direct observational comparisons because both studies likewise include the unresolved broad-plus-narrow H$\alpha$ emission.

However, reproducing the absolute position of the LRDs does not imply that the model reproduces the relative differences between LRDs and LBDs. In particular, our LRDs have a lower median H$\alpha$ equivalent width than LBDs,
$699^{+36}_{-42}$~\AA\ compared with $998^{+51}_{-29}$~\AA, despite exhibiting substantially broader and more luminous H$\alpha$ emission. This combination is difficult to explain through continuum attenuation and orientation alone.

Several limitations should be considered when interpreting these results. First, the sample is dominated by low-resolution NIRSpec/PRISM spectroscopy, in which H$\alpha$ is blended with [N\,II]. The measured flux may therefore include an unresolved [N\,II] contribution, systematically overestimating both $F_{\mathrm{H}\alpha}$ and $L_{\mathrm{H}\alpha}$. Although this affects both populations, any differential bias between LRDs and LBDs remains uncertain. More detailed multi-component modelling could alter the inferred line fluxes and widths, but cannot be applied robustly and homogeneously to most of the present spectra. Nevertheless, the observed separation is large; the median H$\alpha$ luminosities differ by $0.75$ dex, with
$\log_{10}(L_{\mathrm{H}\alpha}/\mathrm{erg\,s^{-1}})=42.63$ for LRDs and $41.88$ for LBDs.

Second, the H$\alpha$ equivalent-width criterion used to construct the common parent sample may affect the comparison between the two populations, particularly near the imposed limits of
$200~\text{\AA}<EW(\mathrm{H}\alpha)<2000~\text{\AA}$. The difference in their median equivalent widths should therefore be interpreted cautiously. However, the same criterion is applied to both populations, and it cannot by itself explain the much larger separation in H$\alpha$ luminosity or line width.

Finally, the DJA combines surveys with heterogeneous selection functions and is not representative of a uniformly selected galaxy population. Its H$\alpha$ parent sample has a median rest-frame equivalent width of $207^{+2}_{-2}$~\AA, broadly consistent with typical high-redshift star-forming galaxies, although biases towards bright, high-redshift, or strong-line sources may remain. Given the broad range of reported LRD properties \citep{Brazzini2026,Geris2026}, the heterogeneous DJA selection may preferentially recover their brighter or more readily detectable members. Nevertheless, our selection identifies 17 LRDs beyond those reported by \citet{deGraaff2025DJA} within the same parent archive, demonstrating that it is not restricted to previously known sources.

\subsection{Luminosity dependence and physical interpretation}

An independent constraint comes from the strong increase in the LRD fraction with H$\alpha$ luminosity. LRDs are rare at the low H$\alpha$-luminosity end but become increasingly dominant towards larger $L_{\mathrm{H}\alpha}$. Within $4<z<7.5$, the cumulative analysis shows that LRDs constitute a statistically robust majority above $\log_{10}(L_{\mathrm{H}\alpha}/\mathrm{erg\,s^{-1}})=42.30$, where the lower bound of the 68\% Wilson interval first exceeds 0.5. By contrast, the LRD fraction shows no clear monotonic evolution with redshift over the range probed, while LRDs remain systematically more H$\alpha$ luminous than LBDs.

This H$\alpha$ luminosity dependence is difficult to reconcile with a simple orientation-based scenario in which the probability of observing an intrinsically similar source as an LRD is independent of its line luminosity. The offset also persists after imposing the same broad-line criterion, ${\rm FWHM}(\mathrm{H}\alpha)>2000~\mathrm{km\,s^{-1}}$, on both populations and therefore cannot be attributed solely to their different line-width distributions. Instead, the increasing incidence of LRDs towards high $L_{\mathrm{H}\alpha}$ suggests that the LRD phenomenon depends on an intrinsic property such as luminosity, accretion state, gas column density, or luminosity-dependent obscuration.

Interestingly, our H$\alpha$ luminosity and line-width measurements agree rather well with the latest predictions of the gas-cocoon scenario developed by \citet{Sneppen2026_cocoonmodel,Sneppen2026_LBDLRDs}, in which LBDs and LRDs form a sequence driven by increasing gas column density. The model predicts LRDs to be approximately $7.5$ times more luminous in H$\alpha$ than LBDs and to exhibit broader lines, with mean FWHM values of $2080~\mathrm{km\,s^{-1}}$ and $1530~\mathrm{km\,s^{-1}}$, respectively. These predictions are close to our observed average H$\alpha$ luminosity ratio of $\simeq5.6$ and median FWHM values of $2319^{+71}_{-64}$ and $1424^{+56}_{-43}~\mathrm{km\,s^{-1}}$ for LRDs and LBDs respectively. 

\begin{figure}
    \includegraphics[width=\columnwidth]{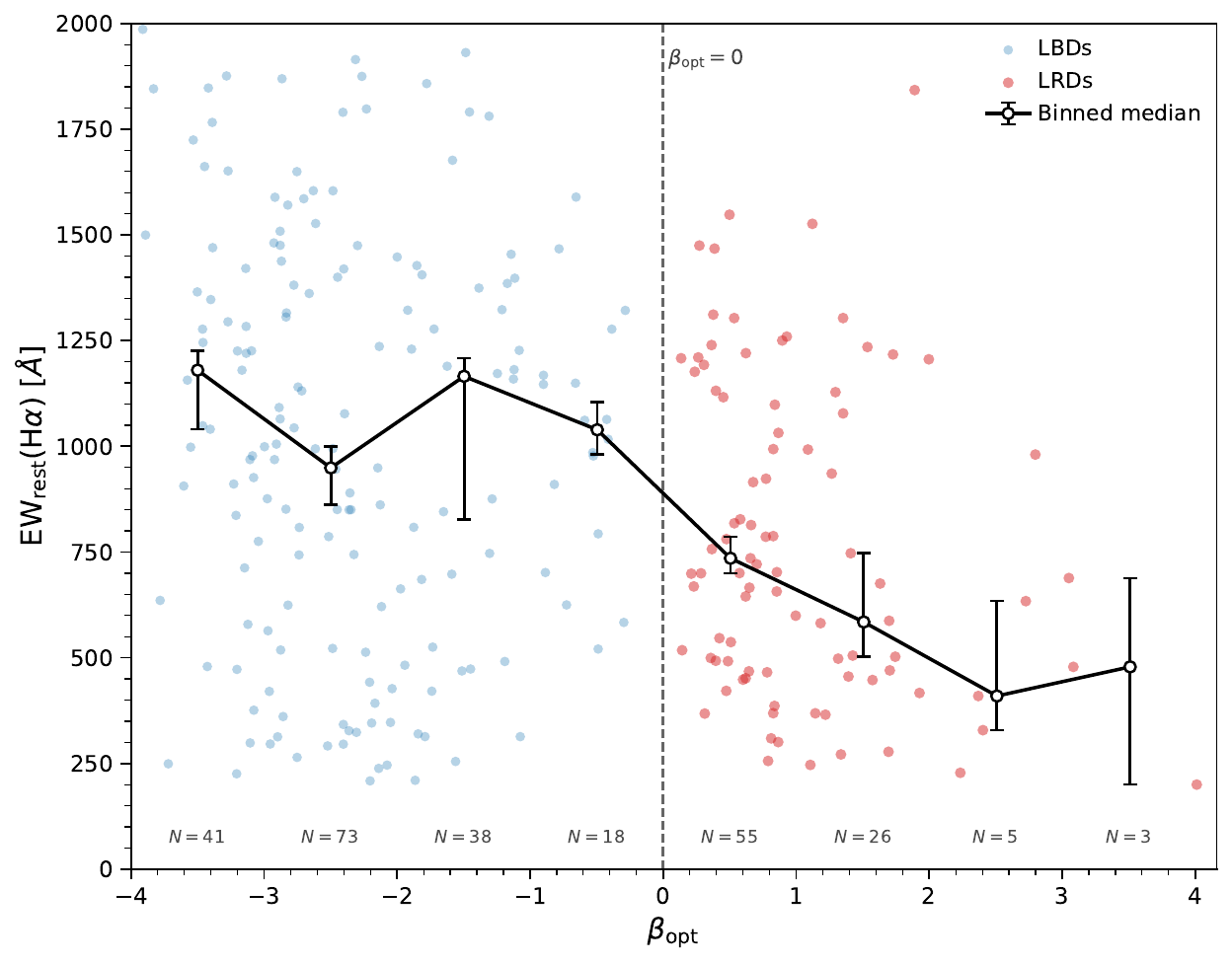}
  \caption{Rest-frame H$\alpha$ equivalent width as a function of the rest-frame optical continuum slope, $\beta_{\rm opt}$, for LBDs (blue circles) and LRDs (red circles). The vertical dashed line marks the adopted $\beta_{\rm opt}=0$ boundary between the two populations. Black points and error bars show the median $\mathrm{EW}_{\rm rest}(\mathrm{H}\alpha)$ and the 16th--84th percentile bootstrap uncertainties in bins of $\beta_{\rm opt}$. The median H$\alpha$ equivalent width is broadly high and relatively stable over the LBD regime, with values of $\sim950$--$1180\,\mathrm{\mathring{A}}$ while $\beta_{\rm opt}<0$. However, it systematically decreases towards redder optical slopes in the LRD regime, from $735^{+51}_{-36}\,\mathrm{\mathring{A}}$ at $0<\beta_{\rm opt}<1$ to $410^{+224}_{-81}\,\mathrm{\mathring{A}}$ at $2<\beta_{\rm opt}<3$. This trend suggests that the H$\alpha$ equivalent width decreases as the optical continuum becomes redder, particularly among LRDs, qualitatively consistent with gas-cocoon models in which an increasing fraction of the H$\alpha$ emission is reprocessed into continuum emission at high gas column densities. The most extreme red bins contain only a small number of sources and should therefore be interpreted with caution.}
    \label{fig: EWHalpha_betaopt}
\end{figure}

The behaviour of $EW(\mathrm{H}\alpha)$ across optical slope provides further support for this interpretation. As shown in Fig.\,\ref{fig: EWHalpha_betaopt},  the median equivalent width remains broadly high and relatively stable across the LBD regime ($\beta_{\rm opt}<0$), with values ranging from $949^{+50}_{-87}$~\AA\ to $1180^{+46}_{-140}$~\AA, before declining towards redder optical slopes in the LRD regime, from $735^{+51}_{-36}$~\AA\ at $0<\beta{\rm opt}<1$ to $585^{+163}_{-82}$~\AA\ at $1<\beta{\rm opt}<2$ and $410^{+224}_{-81}$~\AA\ at $2<\beta{\rm opt}<3$. This decline is qualitatively consistent with the behaviour predicted by the gas-cocoon models of \citet{Sneppen2026_cocoonmodel,Sneppen2026_LBDLRDs}, in which $EW(\mathrm{H}\alpha)$ reaches a maximum at intermediate column densities, approaching the transition between the LBD and LRD regimes, before decreasing again at larger gas column densities as an increasing fraction of the line emission is reprocessed into continuum emission. The lower equivalent widths of the reddest LRDs, despite their substantially larger H$\alpha$ luminosities, are therefore consistent with the expectation that these most extreme/red LRDs occupy the high-column-density regime of the model. In this picture, the systematically higher FWHM of the H$\alpha$ lines in the LRDs as compared to LBDs is also a consequence of increasing column density (rather than projected gas velocities) with an increasing fraction of the H$\alpha$ emission scattered into the observed exponential wings.

Viewed together, our results are inconsistent with a purely orientation-based interpretation of the LRD--LBD connection such as that proposed by \citet{Madau2026}, at least for some fraction of the LRD population. The model unifying LBDs and LRDs purely by orientation approximately reproduces the absolute Balmer decrement and H$\alpha$ equivalent width of the LRDs in our sample, and can, to some extent, explain the increased H$\alpha$ FWHM displayed by LRDs relative to LBDs. However, it cannot readily explain the substantially more luminous typical H$\alpha$ line emission produced by LRDs, the trends in EW with $\beta_{\rm opt}$ across the LBD+LRD population, nor the systematically lower [O\,III]/H$\beta$ ratios displayed by the LRDs. Orientation and dust attenuation may contribute to the red optical continua and large Balmer decrements of LRDs, but the observations require at least one additional luminosity- or accretion-dependent physical parameter. Taken at face value, our results favour a scenario analogous to the model developed by \citet{Sneppen2026_cocoonmodel}, in which LBDs are linked to LRDs via an increasing hydrogen column density in the cocoon surrounding the central engine \citep[see also][]{Inayoshi2025, deGraaff2025DJA, Sneppen2026_LBDLRDs, Rusakov2026}.

\section{Summary and Conclusions}
\label{sec:summary_conclusions}

In this work, we have investigated whether Little Red Dots (LRDs) and Little Blue Dots (LBDs) can be understood as a single compact AGN population observed at different orientations, or whether their properties require additional physical differences. We constructed a homogeneous spectroscopic sample from the DJA archive, selecting on high H$\alpha$ equivalent width, UV continuum slope, and compactness, independently of the optical continuum shape. We subsequently classified the sources according to their position in the $\beta_{\rm UV}$--$\beta_{\rm opt}$ plane and compared their continuum properties, H$\alpha$ emission, and Balmer decrements. Our final sample spans the redshift range $1 \lesssim z \lesssim 7.5$, with the LRD and LBD sub-samples displaying very similar redshift distributions (with median redshifts $z_{\rm med}=4.53^{+0.30}_{-0.30}$ for LRDs and $z{\rm med}=4.80^{+0.27}_{-0.21}$ for LBDs). Our main conclusions can be summarised as follows: 

\begin{enumerate}

    \item Our final sample comprises 280 compact, UV-blue, strong-H$\alpha$ sources: 89 LRDs and 191 LBDs. Of the LRDs, 72 were previously reported by \citet{deGraaff2025RUBIES}, while 17 are newly identified in this work. The two populations are strongly separated in optical continuum slope, with median values of $\beta_{\rm opt}=0.83^{+0.03}_{-0.05}$ for LRDs and $\beta_{\rm opt}=-2.63^{+0.15}_{-0.12}$ for LBDs. The newly identified sources extend the known LRD population towards less extreme UV--optical continuum breaks.

    \item LRDs have systematically broader H$\alpha$ profiles than LBDs, with median FWHM values of $2319^{+71}_{-64}~\mathrm{km\,s^{-1}}$ and $1424^{+56}_{-43}~\mathrm{km\,s^{-1}}$, respectively. The distributions are statistically distinct, indicating that the two populations likely differ in gas kinematics. 

    \item The strongest distinction is found in H$\alpha$ luminosity. LRDs have a median $\log_{10}(L_{\rm H\alpha}/\mathrm{erg\,s^{-1}})=42.63^{+0.09}_{-0.09}$, compared with $41.88^{+0.04}_{-0.01}$ for LBDs. The offset corresponds to a factor $\sim6$ in luminosity and persists after imposing the same broad-line criterion, $\mathrm{FWHM}(\mathrm{H}\alpha)>2000~\mathrm{km\,s^{-1}}$, on both populations. It therefore cannot be explained solely by their different FWHM distributions or by their redshift coverage.

    \item LRDs have a moderately lower median EW than LBDs, $699^{+36}_{-42}$~\AA\ compared with $998^{+51}_{-29}$~\AA. Because H$\alpha$ EW enters the parent-sample selection, this difference should be interpreted cautiously. Nevertheless, the combination of lower EW and substantially higher H$\alpha$ luminosity shows that the enhanced line luminosities of LRDs are not accompanied by correspondingly larger line-to-continuum ratios. Moreover, the H$\alpha$ EW decreases systematically towards the reddest optical slopes, reaching $544^{+94}_{-55}$~\AA\ in the reddest bin, qualitatively consistent with gas-cocoon models in which increasing gas column densities cause an increasing fraction of the line emission to be reprocessed into the optical continuum. 

    \item The fraction of LRDs increases strongly with H$\alpha$ luminosity, with LRDs becoming increasingly dominant towards the high-luminosity end. This trend disfavours a purely orientation-driven scenario in which the probability of observing a source as an LRD is independent of its intrinsic line luminosity. Instead, LRDs preferentially occupy the high-$L_{\rm H\alpha}$ regime of the compact broad-line population.

    \item LRDs are strongly concentrated at large Balmer decrements, with a median $\log_{10}(F_{\rm H\alpha}/F_{\rm H\beta})=1.01\pm0.03$, compared with $0.47\pm0.01$ for LBDs. The LRD fraction increases sharply with Balmer decrement, and H$\alpha$ EW of LRDs are clearly offset from the low-redshift broad-line QSO population. This indicates that the Balmer-emitting gas in LRDs is subject to substantially different attenuation and radiative-transfer conditions than in LBDs.

    \item LRDs also show systematically lower [O III]~$\lambda5007$/H$\beta$ ratios than LBDs. Because these two lines are closely spaced in wavelength, their ratio is only weakly affected by dust attenuation, suggesting that the difference is unlikely to arise from foreground reddening alone and may instead reflect differences in ionisation conditions, gas density, metallicity, or the visibility of the narrow-line region.

\end{enumerate}

Overall, our results are inconsistent with a simple orientation-only interpretation of the LRD-LBD connection. Although viewing geometry and dust attenuation likely contribute to the red optical continua and large Balmer decrements of LRDs, and can potentially explain their broader H$\alpha$ lines, they do not readily explain the substantially more luminous H$\alpha$ emission produced by LRDs, their lower [O III]/H$\beta$ ratios, or the decline of H$\alpha$ EW towards redder optical slopes. Our results therefore favour a scenario in which orientation may contribute to the observed diversity, but in which increasing gas column density around the central engine provides a plausible mechanism for linking LBDs and LRDs. 

Rather than forming two sharply separated populations, LBDs and LRDs may represent different regimes within a broader population of compact, high-redshift accreting sources, with increasing gas column density providing a possible link between their observed properties.

\section*{Data availability}

All catalogues and software products generated for this work will be made publicly available in a GitHub repository upon publication of this paper. The repository will include the final catalogues, the Jupyter notebooks used for the analysis, and the Python scripts required to reproduce the results and figures presented in this work.

\noindent\textit{Software:} This work made use of the Python packages
\texttt{Astropy} \citep{Astropy2022}, \texttt{NumPy}
\citep{Harris2020}, \texttt{SciPy} \citep{Virtanen2020},
\texttt{Pandas} \citep{McKinney2010}, \texttt{Matplotlib}
\citep{Hunter2007}, and \texttt{Photutils} \citep{Bradley2026},
as well as Jupyter notebooks for data analysis and visualisation.



\bibliographystyle{mnras}
\bibliography{LRDspaper_bibliography} 

\section{Acknowledgements}

LB and JSD acknowledge the support of the Royal Society through
the award of a Royal Society University Research Professorship to
JSD. LB and MSL thank Pedro Rodríguez Pascual (XMM-Newton Science
Operations Centre, ESAC) for his assistance with the X-ray data. 
DHC acknowledges support from the Basque Government through the Programa Predoctoral de Formación de Personal Investigador No Doctor del departamento de Ciencia, Universidades e Innovación del Gobierno Vasco

This research has made use of data from the JWST/NIRSpec spectroscopic
programmes contributing to the final sample, including JADES,
CAPERS, RUBIES, UNCOVER, CEERS, NEXUS, CANUCS, GTO-WIDE, AURORA,
GLASS-JWST, DeepDive, and other General Observer and Director's
Discretionary Time programmes. We acknowledge the principal
investigators, teams, and staff involved in obtaining and processing
these observations.

This work is based on observations made with the NASA/ESA/CSA
James Webb Space Telescope. The data products presented herein were
retrieved from the Dawn JWST Archive (DJA), an initiative of the
Cosmic Dawn Center (DAWN), which is funded by the Danish National
Research Foundation under grant DNRF140.

This research has made use of data obtained from the 5XMM
serendipitous source catalogue compiled by the XMM-Newton Survey
Science Centre, the XMM2ATHENA project, and in collaboration with
the XMM-Newton Science Operations Centre. This research has also
made use of data obtained from the Chandra Source Catalogue,
provided by the Chandra X-ray Center (CXC) as part of the Chandra
Data Archive. 



\appendix
\section{Additional sample properties}
\label{sec :additional_properties}

This appendix provides supplementary information on the survey composition of our sample and illustrates the effects of the main selection criteria on the H$\alpha$ equivalent-width and optical-slope distributions. 

Table~\ref{tab:survey_summary} lists the contribution of each spectroscopic survey, the field observed and the number of sources to the final sample.

\begin{figure}
    \includegraphics[width=\columnwidth]{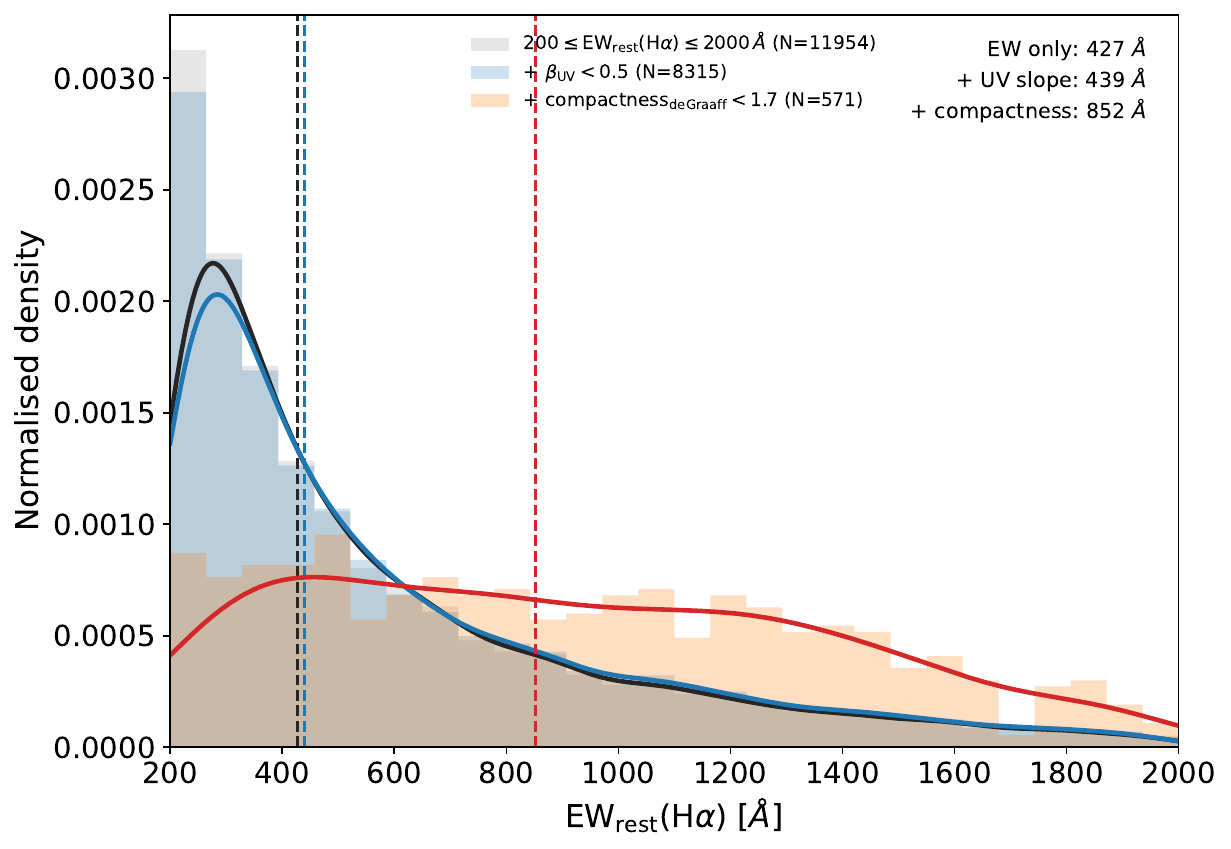}
    \caption{Normalised distributions of the rest-frame H$\alpha$ equivalent width after the sequential application of the selection criteria. The black distribution shows the 11,954 sources satisfying $200 \leq \mathrm{EW}_{\rm rest}(\mathrm{H}\alpha) \leq 2000$\,\AA, while the blue distribution additionally imposes $\beta_{\rm UV}<-0.5$, leaving 8,315 sources. The red distribution corresponds to the final sample after applying the compactness criterion, comprising 571 sources. Dashed vertical lines indicate the respective median equivalent widths. The UV-slope cut has little effect on the distribution, shifting the median from 427 to 439\,\AA. By contrast, the compactness criterion substantially reduces the sample and shifts the median to 852\,\AA.}
    \label{fig:EW_compactness_criteria}
\end{figure}

Figure~\ref{fig:EW_compactness_criteria} shows that the UV-slope criterion has little effect on the H$\alpha$ equivalent-width distribution, whereas the compactness cut preferentially selects sources with larger equivalent widths. 

\begin{figure}
    \includegraphics[width=\columnwidth]{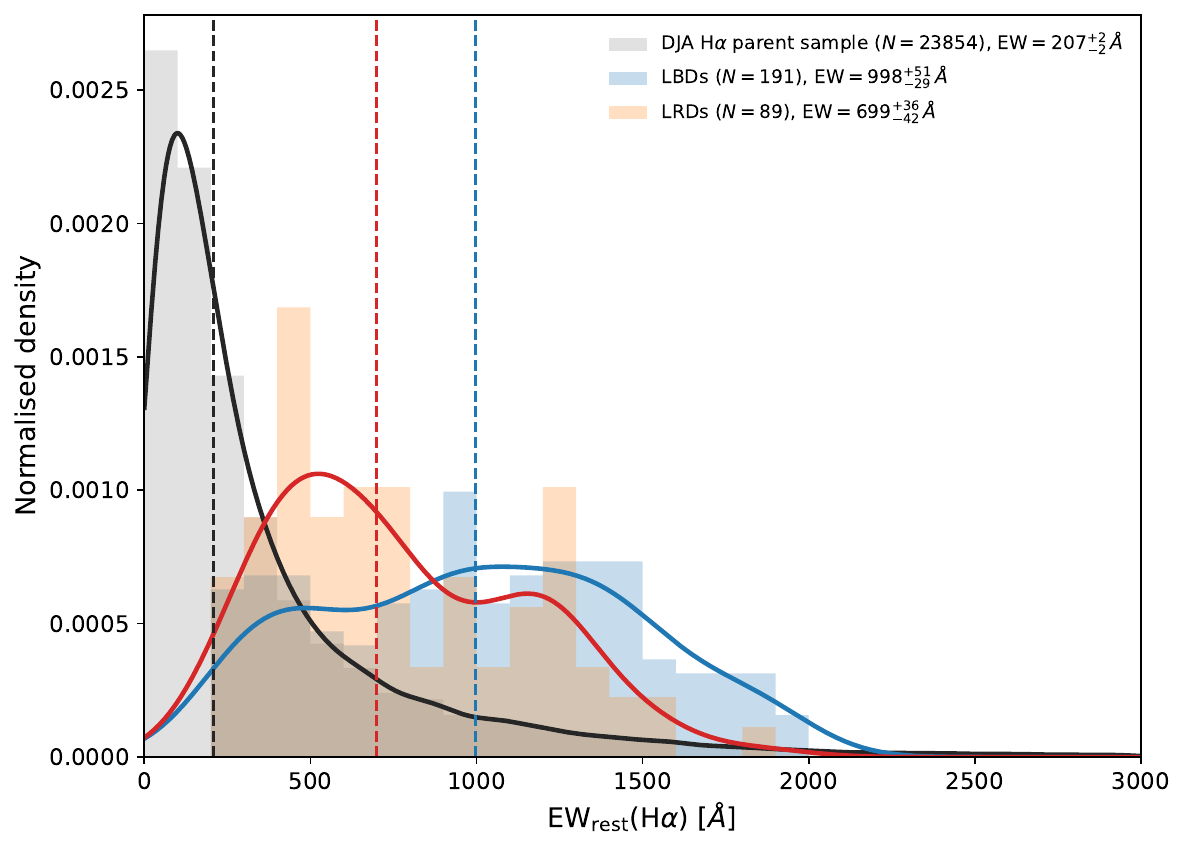}
    \caption{ Normalised distributions of the rest-frame H$\alpha$ equivalent width for the DJA H$\alpha$ parent sample (black), LBDs (blue), and LRDs (red). The parent sample comprises 23,854 sources with successful H$\alpha$ fits and valid equivalent-width measurements within the displayed range, although its distribution may reflect the heterogeneous and predominantly high-redshift nature of the DJA archive. Dashed vertical lines indicate the median equivalent widths, with bootstrap uncertainties reported in the legend. Both selected populations are strongly shifted towards larger equivalent widths relative to the parent sample, whose median is $207^{+2}_{-2}$\,\AA. LBDs exhibit a larger median equivalent width than LRDs, with $998^{+51}_{-29}$\,\AA\ and $699^{+36}_{-42}$\,\AA, respectively. } 
    \label{fig:EW_fullDJAparent_vs_LRDs_LBDs}
\end{figure}

Figure~\ref{fig:EW_fullDJAparent_vs_LRDs_LBDs} compares the selected populations with the DJA parent sample. Both LRDs and LBDs exhibit substantially larger H$\alpha$ equivalent widths than the parent population, with LBDs reaching the largest median value. The parent sample comprises approximately 24,000 sources with measurable H$\alpha$. 
Its distribution peaks around $\mathrm{EW}_{\rm rest}({H}\alpha)\sim200,\AA$, broadly consistent with the values expected for typical high-redshift star-forming galaxies. 
Nevertheless, given the heterogeneous selection of the surveys included in the DJA, the parent sample should not be regarded as fully representative of the general galaxy population. However, it broadly covers the H$\alpha$ equivalent-width range expected for typical galaxies.

\begin{figure}
    \includegraphics[width=\columnwidth]{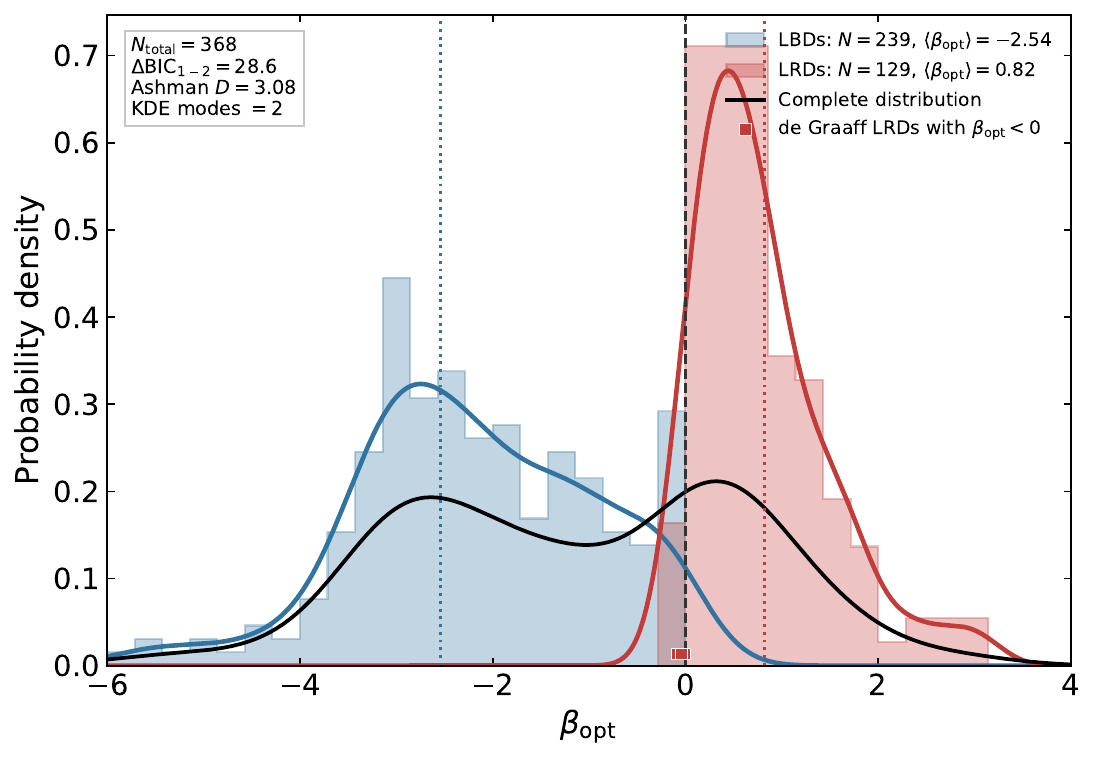}
    \caption{Distribution of the optical continuum slope, $\beta_{\rm opt}$, after applying all selection criteria except the Monte Carlo significance requirement on $\beta_{\rm opt}$. The normalised histograms and kernel-density estimates show LBDs (blue) and LRDs (red), while the black curve represents the complete sample. The vertical dotted lines mark the population means and the dashed line indicates $\beta_{\rm opt}=0$; red squares identify previously known de Graaff LRDs with $\beta_{\rm opt}<0$. A two-component Gaussian-mixture model is strongly preferred over a single Gaussian ($\Delta{\rm BIC}_{1-2}=28.6$), with well-separated components (Ashman’s $D=3.08$), while the KDE independently recovers two modes. These results indicate that the $\beta_{\rm opt}$ distribution is intrinsically bimodal before imposing its significance criterion.}   
    \label{fig:bimodality_before_betaoptcleaning}
\end{figure}

Finally, Fig.~\ref{fig:bimodality_before_betaoptcleaning} examines the $\beta_{\rm opt}$ distribution after applying all selection criteria except the Monte Carlo requirement that its sign be recovered with $\geq95$ per cent confidence. We fitted the 356 sources within $-6\leq\beta_{\rm opt}\leq4$ with one- and two-component Gaussian-mixture models using \texttt{GaussianMixture} from \texttt{scikit-learn}. The two-component model is strongly preferred over a single Gaussian ($\Delta{\rm BIC}=28.65$), with components centred at $\beta_{\rm opt}=-2.53$ and $0.44$. These components are clearly separated (Ashman’s $D=3.08$), and a kernel-density estimate independently recovers peaks at $\beta_{\rm opt}\simeq-2.64$ and $0.32$. Thus, the bimodality is already present before imposing the Monte Carlo significance criterion.

\begin{table}
  \caption{Spectroscopic surveys contributing to the final sample of 280 sources. The table lists the survey, field, and number of sources contributed by each programme. The sample combines spectroscopic observations from 19 surveys and programmes, with CAPERS and RUBIES providing the largest contributions.}
\label{tab:survey_summary}  
     \begin{tabular}{llr}
        \hline
        Survey & Field & No. \\
        \hline
        CAPERS
        \citep{Dickinson2024}
        & UDS/COSMOS & 65 \\

        JADES
        \citep{Eisenstein2023}
        & GOODS-N/S & 54 \\

        RUBIES
        \citep{deGraaff2025RUBIES}
        & EGS/UDS & 43 \\

        UNCOVER
        \citep{Bezanson2024}
        & Abell~2744 & 24 \\

        CEERS
        \citep{Finkelstein2025}
        & EGS & 18 \\

        NEXUS
        \citep{Shen2024}
        & --- & 18 \\

        GTO-WIDE
        \citep{Maseda2024}
        & GOODS-S/EGS/UDS & 16 \\

        CANUCS
        \citep{Sarrouh2025}
        & Cluster fields & 12 \\

        DDT-6585
        & COSMOS & 7 \\

        GO-4106
        & EGS & 4 \\

        GO-2073
        & Quasar fields & 4 \\

        MoM
        \citep{Naidu2025}
        & --- & 4 \\

        GO-1433
        \citep{Coe2023}
        & MACS0647 & 3 \\

        GO-2198
        \citep{Barrufet2025}
        & GOODS-S & 3 \\

        NIRSpec WIDE
        \citep{Maseda2024}
        & COSMOS & 1 \\

        AURORA
        \citep{Shapley2025}
        & COSMOS/GOODS-N & 1 \\

        GO-2767
        & RXJ2129 & 1 \\

        GLASS-JWST
        \citep{Treu2022}
        & Abell~2744 & 1 \\

        DeepDive
        \citep{Valentino2025}
        & EGS/UDS & 1 \\

        \hline
        \textbf{Total} & & \textbf{280} \\
        \hline
  
    \end{tabular}
\end{table}

\bsp	
\label{lastpage}
\end{document}